\documentclass[aoas,preprint]{imsart}

\RequirePackage[OT1]{fontenc}
\RequirePackage{amsthm,amsmath}
\usepackage{amsfonts}
\RequirePackage[authoryear]{natbib}
\RequirePackage[colorlinks,citecolor=blue,urlcolor=blue]{hyperref}
\usepackage{graphicx}
\usepackage{bbm}
\usepackage{algorithm}
\usepackage[noend]{algpseudocode}
\usepackage{caption}

\makeatletter
\def\BState{\State\hskip-\ALG@thistlm}
\makeatother

\newcommand{\0}{{\mathbf{0}}}

\newcommand{\m}{{\mathbf{m}}}

\newcommand{\y}{{\mathbf{y}}}

\newcommand{\A}{{\mathbf{A}}}

\newcommand{\X}{{\mathbf{X}}}

\newcommand{\T}{{\rm{T}}}

\newcommand{\bbeta}{\boldsymbol{\beta}}

\newcommand{\btau}{\boldsymbol{\tau}}
\newcommand{\btheta}{\boldsymbol{\theta}}

\arxiv{arXiv:0000.0000}

\startlocaldefs
\numberwithin{equation}{section}
\theoremstyle{plain}

\endlocaldefs

\begin{document}

\begin{frontmatter}
\title{Power Weighted Densities for Time Series Data}
\runtitle{Power Weighted Densities for Time Series Data}

\begin{aug}
\author{\fnms{Daniel} \snm{McCarthy}} \and
\author{\fnms{Shane T.} \snm{Jensen}}

\affiliation{University of Pennsylvania and University of Pennsylvania}

\address{Daniel McCarthy\\
Department of Statistics \\
The Wharton School \\ 
University of Pennsylvania \\
400 Jon M. Huntsman Hall \\
3730 Walnut Street\\
Philadelphia, PA 19104\\
Email: \href{mailto:danielmc@wharton.upenn.edu}{danielmc@wharton.upenn.edu}}

\address{Shane T. Jensen\\
Department of Statistics \\
The Wharton School \\ 
University of Pennsylvania \\
400 Jon M. Huntsman Hall \\
3730 Walnut Street\\
Philadelphia, PA 19104\\
Email: \href{mailto:stjensen@wharton.upenn.edu}{stjensen@wharton.upenn.edu}}

\end{aug}

\begin{abstract}
While time series prediction is an important, actively studied problem, the predictive accuracy of time series models is complicated by non-stationarity.  We develop a fast and effective approach to allow for non-stationarity in the parameters of a chosen time series model.  In our power-weighted density (PWD) approach, observations in the distant past are down-weighted in the likelihood function relative to more recent observations, while still giving the practitioner control over the choice of data model.   One of the most popular non-stationary techniques in the academic finance community, rolling window estimation, is a special case of our PWD approach.  Our PWD framework is a simpler alternative compared to popular state-space methods that explicitly model the evolution of an underlying state vector.   We demonstrate the benefits of our PWD approach in terms of predictive performance compared to both stationary models and alternative non-stationary methods.  In a financial application to thirty industry portfolios, our PWD method has a significantly favorable predictive performance and draws a number of substantive conclusions about the evolution of the coefficients and the importance of market factors over time.
\end{abstract}

\begin{keyword}[class=MSC]
\kwd[Primary ]{37M10}
\kwd{}
\kwd[Secondary ]{62P05}
\end{keyword}

\begin{keyword}
\kwd{Time series analysis; power prior; forecasting; finance}
\end{keyword}

\end{frontmatter}

%%%%%%%%%%%%%%%%%%%%%%%%%%%%%%%%%%%%%%%%%%%%%%%%%%
\section{Introduction and Motivation} \label{intro}
%%%%%%%%%%%%%%%%%%%%%%%%%%%%%%%%%%%%%%%%%%%%%%%%%%

An increasingly prominent area of statistical application is the modeling of data that is ordered over time, either as a single time series or multiple time series, with the goal being the prediction of future time series data.  It is often unrealistic to assume stationarity, whereby the underlying parameters of the chosen model are constant over time. Rather, it may be preferred to allow the parameters of the model to evolve over time, which complicates modeling efforts.  We propose a general methodology which may be used to improve the predictive accuracy of time series models by addressing possible non-stationarity in model parameters.

In the time series application that we focus upon, the issue of non-stationarity is particularly acute -- estimation of the sensitivity of stock returns to market factors.  \citet{fama1993common} introduced the popular three factor model in asset pricing, which relates the return on stock portfolios to their valuation, size and sensitivity to the overall market.   Specifically, the returns $y_{j,t}$ of a stock portfolio $j$ at time $t$ were modeled as a linear function of three factors,
\begin{eqnarray}  \label{ff3factor}
y_{j,t} = \alpha_{j} \, + \, \beta^m_{j} \, \cdot \, m_{t} \, + \, \beta^s_{j}  \, \cdot \,  s_{t} \, + \, \beta^v_{j}  \, \cdot \,  v_{t} \, + \, \epsilon_{j,t}  \label{3factor-intro}
\end{eqnarray}
where $m_{t}$ represents excess return on the market portfolio (`MKT'), $s_{t}$ represents excess return of small capitalization stocks over large capitalization stocks (`SMB'), $v_{t}$ represents the excess return of value stocks over growth stocks (`HML') and $\epsilon_{j,t}$ is a noise term.  Since then, hundreds of papers have been written trying to explain cross-sectional heterogeneity in asset price returns through the inclusion of additional factors.  The overarching goal of this literature is to explain variation in returns across stocks through a relatively small number of market factors, which is equivalent to predicting stock returns using contemporaneous predictors in a time series regression. 

Time series regression problems like this one are notoriously challenging because the parameters of the regression model are unlikely to be stationary over time.  The sensitivity of parameters should be allowed to evolve over time (e.g. $\beta^m_{j,t}$ rather than $\beta^m_{j}$, $\beta^s_{j,t}$ rather than $\beta^s_{j}$, etc., in Equation \ref{ff3factor}).  The question here, and in many other applied settings, is how to address potential non-stationarity in the parameters of a chosen model?  Throughout the remainder of this paper, we will use the term `non-stationarity' to mean that the parameters of the true underlying process generating the observed data are potentially varying over time. 

Our methodological objective is to produce the best possible {\it predictions at the next time point}, conditional upon the model the practitioner has chosen.  If we are unwilling to assume stationarity over time for the model parameters, the consequence is that not all historical data will be equally relevant to the prediction of future outcomes.  With prediction as our ultimate goal, we will propose statistical methodology for a principled differential weighting of historical data that is simple and efficient relative to traditional methods that focus on estimation of the underlying parameter evolution.  While this paper explores an application to market factor sensitivies, our {\it power-weighted densities} (PWD) approach can be applied to any time series setting where the underlying data generating process is believed to be non-stationary over time.

Financial data are an interesting case study for time series methods as many assets have been tracked for a relatively long time period.   In this paper, we will model the monthly returns of 30 industry portfolios (\href{http://mba.tuck.dartmouth.edu/pages/faculty/ken.french/data_library.html#Research}{Kenneth French}).  The time series begin in July 1926 and end in December 2014, which gives us 1062 time points for each of 30 stock portfolios.

However, the long length of these time series is deceptive due to non-stationarity in the underlying data generating process.  Acknowledging this non-stationarity, practitioners usually employ some sort of data truncation, ignoring data which is `old enough' under the assumption that market conditions make data prior to that point irrelevant or even harmful to the predictive accuracy of their model.   

In the finance literature, non-stationary is usually addressed by estimating asset models using {\it rolling windows}, i.e.  assuming a stationary model in a fixed window of data closest to the current time point.  The key question is how long should one make the rolling window length?  \citet{petkova2005value} chose a 5 year rolling window while \citet{fama1993common} chose a 30 year rolling window.   As part of their comparison of equity risk premium theories, \citet{welch2008comprehensive} use an expanding rolling window: at each time point $t$, they use all data up to and including time point $t$.   While {\it explicit} data truncation via rolling window estimation is very frequently employed, {\it implicit} data truncation may be at least as prevalent, by pre-specifying the date range over which analysis will be performed.  We seek a more principled approach to addressing non-stationarity in time series without relying on {\it ad hoc} decisions of how to truncate the data.  

In the general approach to time-ordered data, a practitioner has chosen a model $p(y_t | \btheta)$ that links the observed data  $\y_{1:\T} = (y_1, \cdots, y_\T) $  to underlying parameters $ \btheta$.   The practitioner may also have prior beliefs summarized in the prior distribution $p_0(\btheta)$.  The simplest Bayesian approach to modeling $\y_{1:\T}$ would be to assume that $\btheta$ is stationary over time and estimate the posterior distribution assuming the observed $y_t$'s are exchangeable,
\begin{equation} \label{cb-eqn}
p (\btheta | \y_{1:\T}) \propto \prod_{t=1}^{\T} p(y_t | \btheta) \, p_0(\btheta).
\end{equation}
However, as we discussed above, stationarity is not always a reasonable assumption and so we need to allow for the  underlying parameters of the model to evolve over time, i.e. $\btheta_{1:\T} = (\btheta_1, \btheta_2, \cdots, \btheta_{\T-1}, \btheta_\T) $.  

A standard Bayesian approach to non-stationarity specifies an additional level of the model for this parameter evolution (i.e. $\btheta_t$ given $\btheta_{1:t-1}$) such as the dynamic state-space model \cite[]{west1998bayesian}.    In addition to these extra modeling decisions, implementation is much more involved since the posterior distribution for an entire time-varying series of parameters, 
\begin{equation} \label{dlm}
p (\btheta_{1:\T} | \y_{1:\T}) \propto \prod_{t=1}^{\T} p(y_t | \btheta_t) p( \btheta_t |  \btheta_{1:t-1}) \, p_0(\btheta).
\end{equation}
must be estimated.     Under the simplifying assumption of normality, \citet{CarKoh94} outline a Markov Chain Monte Carlo implementation for estimating the posterior distribution of a dynamic state-space model.  More recent work has offered implementations for more complicated dynamic state-space models \citep{paez2013hierarchical}.   However, all of these modeling approaches are inherently complicated (and usually computationally intensive) because the entire time-varying parameter vector $\btheta_{1:\T}$ must be estimated.  

In contrast, we propose an alternative {\it power-weighted densities} (PWD) approach that avoids the direct specification of an evolution model for the parameter vector $\btheta_{1:\T}$.  We leave the practitioners' chosen model as is, but differentially weight the contribution of individual observations to the likelihood function, so that more recent observations are more informative in the posterior distribution of the parameters at the current time point.   Specifically, as we will see in Section~\ref{methodology}, 
\begin{equation} \label{pp-eqn-intro}
p_\alpha(\btheta_\T | \y_{1:\T}) \propto p_0(\btheta_\T) \prod_{i=0}^{\T-1} p(y_{\T-i} | \btheta_\T)^{\alpha_i}, \ \ \alpha \in [0,1],
\end{equation}
where $\alpha_i$ are weights placed on the lagged observations $y_{\T-i}$ away from the current time point $\T$.  These weights are estimated from the data in order to optimize the one step ahead predictive likelihood of the observed data.

Our PWD approach leaves intact the basic form of the model, $p(y_t | \btheta)$ and $p_0(\btheta)$, which makes our approach complementary to whatever data model is preferred by the practitioner.    In contrast with the dynamic state-space model (\ref{dlm}), our PWD approach does not require estimation of the entire parameter vector $\btheta_{1:\T}$ in order to infer the posterior distribution of the terminal time-point or to make predictions of future time points, which is the primary objective of our study.

As we will see in Section~\ref{methodology}, rolling windows correspond to a specific set of lag-dependent PWD weights.  While rolling windows also leave the choice of the data model up to the practitioner, we will see that the performance of rolling window approaches can be erratic in practice.  In contrast, our PWD approach avoids the pre-specification of a fixed window length by differentially down-weighting {\it all} previous observations to optimize the predictive likelihood of the observed data.  We present the details of our general power-weighted densities approach to time series data and compare our approach to state space models and other time series methods in in Section \ref{methodology}.  

Our financial application consists of time series for 30 separate stock portfolios, which motivates extending our PWD approach to a hierarchical linear regression model in Section~\ref{hier-lin}.  This extension permits sharing of information between the \citet{fama1993common} three-factor models (\ref{3factor-intro}) for each stock portfolio while addressing non-stationarity within each stock portfolio time series.   A hierarchical model is motivated by the central tendency of the market beta for a large number of stocks, often referred to as `beta decay' by financial practitioners.   

As much has been written about model uncertainty in stock return prediction, we will also incorporate uncertainty about our model choices by outlining a Bayesian Model Averaging (`BMA') extension of our PWD approach in Section \ref{bma}.  Our general PWD methodology for time series will be made available via a R package on CRAN.    

In Section~\ref{sim-eval}, we compare the operating characteristics of our PWD approach for hierarchical linear regression to alternative methods in synthetic data settings that mimic aspects of our financial data.   In Section~\ref{stock-mkt-ind}, we apply our PWD approach to hierarchical linear regression to the monthly returns of 30 industry portfolios (\href{http://mba.tuck.dartmouth.edu/pages/faculty/ken.french/data_library.html#Research}{Kenneth French}).   In both real and synthetic data, our PWD approach performs significantly better in terms of predictive accuracy than models that assume stationarity in the underlying parameters, as well as competing non-stationary approaches such as dynamic state-space models and rolling windows.  We will also demonstrate the computational convenience of our PWD approach.

There are a number of substantive implications of our results for financial practitioners.  First, our results suggest a considerable amount of variation over time in the sensitivity of industry portfolio returns, particularly in the time periods around 1960 and 2000.  Second, we observe a `self-fulfilling prophecy' effect: the publication/acceptance of the importance of a market factor is followed by an increase in the importance of that market factor for prediction.  

%%%%%%%%%%%%%%%%%%%%%%%%%%%%%%%%%%%%%%%%%%%%%%%%%%
\section{Power Weighted Densities for Time Series Data} \label{methodology}
%%%%%%%%%%%%%%%%%%%%%%%%%%%%%%%%%%%%%%%%%%%%%%%%%%

The idea of differentially weighting historical data has been explored previously.  \citet{ibrahim2000power} introduced ``power priors" as a way to integrate historical data with more recent data.  Denoting the historical data by $H$, current data by $\y$, parameters of interest by $\btheta$ and a fixed power $\alpha \in [0,1]$, the posterior distribution from their power prior model is
\begin{equation}
p(\btheta | \y, H, \alpha) \propto p(\y|\btheta) p(H|\btheta)^\alpha p(\btheta) \label{classical-powprior}
\end{equation}
By setting $\alpha = 1$, the historical data is exchangeable with the current data, while $\alpha = 0$ implies the historical data is not used at all.  Power priors have been applied in several clinical and epidemiological studies, including \citet{berry1996bayesian}, \citet{berry2010bayesian},  \citet{hobbs2011hierarchical} and \citet{tan2002bayesian}. \citet{brian2010bayesian} applied power priors to pediatric quality of care evaluation.   

In this paper, we are extending the power prior idea of \citet{ibrahim2000power} to the modeling of time-ordered observations, 
$$ \y \stackrel{\Delta}{=} \y_{1:\T} = (y_1, y_{2}, \cdots, y_{\T-1}, y_\T) $$
motivated by the assumption that older data may not be as relevant as more recent data when predicting future time series outcomes.   We estimate the posterior distribution for $\btheta_\T$ at terminal time point $\T$ by raising the densities of each observation $y_t$ to a different power, 
\begin{equation} \label{pp-eqn-general}
p_\alpha(\btheta_\T | \y_{1:\T}) \propto p_0(\btheta_\T) \prod_{i=0}^{\T-1} p(y_{\T-i} | \btheta_\T)^{\alpha_i}, \ \ \alpha_i \in [0,1].
\end{equation}
which extends the power prior idea to place a lag-specific weight $\alpha_i$ on each $i$-th lagged historical data point away from the current time point. We still encode any prior beliefs we have regarding $\btheta_\T$ through the prior $p_0(\btheta_\T)$.
 
 The density (\ref{pp-eqn-general}) uniquely minimizes the convex sum of Kullback-Leibler divergences over a $\T$-simplex representing all possible poolings of the historical data (further details in \ref{suppA}).    The popular {\it rolling window} strategy for model estimation in the financial literature corresponds to a special case of our PWD weights, where $\alpha_i = 1$ if $i < \tau$ and $\alpha_i = 0$ otherwise, with stopping time $\tau$ being pre-specified by the practitioner.  

By avoiding the estimation of the entire time-series of underlying parameters  $\btheta_{1:\T}$, our PWD approach should be less computationally intensive than the usual dynamic state-space model, but only if the extra weight parameters $\alpha_i$ can be estimated efficiently.  We simplify this estimation task by imposing additional structure on the weight parameters.  

Throughout this paper, we will restrict our weight parameters to an {\it exponentially-decreasing} function, $\alpha_i = \alpha^i$ of the lag $i$, parameterized by a single weight parameter $\alpha \in [0,1]$.   Under this constraint, our {\it power-weighted densities} (PWD) posterior distribution for $\btheta_\T$ at current time point $\T$ becomes
\begin{equation} \label{pp-eqn}
p_\alpha(\btheta_\T | \y_{1:\T}) \propto p_0(\btheta_\T) \prod_{i=0}^{\T-1} p(y_{\T-i} | \btheta_\T)^{\alpha^i}, \ \ \alpha \in [0,1],
\end{equation}
with a single weight parameter $\alpha$ that will be estimated from the data.   This exponentially-decreasing regime of weights imposes a monotonicity constraint $ \alpha_{i} \geq \alpha_{i+1}$ so that with $\alpha \in [0,1]$, more recent observations (those with smaller lags $i$ away from the current time point) have increased relevance relative to more distant observations.   

%When $\alpha$ is strictly less than 1, more recent observations have increased relevance to our primary goal: the prediction of the next time point, $y_{\T+1}$.   

There are many alternatives to our exponentially-decreasing weight regime, with the most obvious alternative being linearly-decreasing weights - we show in \ref{suppA} that linearly decaying weights also perform well in practice.  The exponentially-decreasing regime has the advantage of leading to simple posterior and posterior predictive distributions when used with exponential family likelihoods.

As an illustrative example, consider a single time series $\y_{1:\T}$ that is normally distributed, $y_t \sim \mathcal{N} (\mu_t, \sigma_t^2)$, with unknown and possibly non-stationary mean $\mu_t$ and variance $\sigma_t^2$.    We employ the prior $p(\mu_t, \sigma_t^2) \propto \sigma_t^{-2}$ suggested by \cite{gelman2003bayesian} (p. 74).   Combining this data and prior model with our exponentially-weighted PWD approach (\ref{pp-eqn}) gives the conditional posterior distribution for the terminal mean, 
\begin{eqnarray}   \label{norm-mu-eqn}
\mu_\T \, | \, \y, \alpha, \sigma_\T^2 \quad \sim \quad \mathcal{N} \Big( \hat{y}_{\alpha,\T} \, , \, \dfrac{\sigma_\T^2}{\T_\alpha} \Big), \label{norm-unknownvar-mean}
\end{eqnarray}
and the marginal posterior distribution for the terminal variance,
\begin{eqnarray}
\sigma_\T^2 \, | \, \y, \alpha \quad \sim \quad  \text{InvGamma} \Big(\dfrac{\T_\alpha - 1}{2} \, , \, \dfrac{\T_\alpha}{2} (\widehat{y_{\alpha,\T}^2} - \hat{y}_{\alpha,T}^2)\Big) \label{norm-unknownvar-var}
\end{eqnarray}
where 
\begin{eqnarray*} \label{PWD-notation}
\T_\alpha =  \sum_{i=0}^{\T-1} \alpha^i \,\, , \quad \hat{y}_{\alpha,\T} =  \dfrac{\sum_{i=0}^{\T-1} \alpha^i y_{\T-i} }{\T_\alpha} \,\, , \quad {\rm and} \quad  \widehat{y_{\alpha,\T}^2} = \dfrac{\sum_{i=0}^{\T-1} \alpha^i y_{\T-i}^2}{\T_\alpha}. 
\end{eqnarray*}
The posterior distribution for $\mu_\T$ is centered at $\hat{y}_{\alpha,\T}$, the exponentially weighted moving average (EWMA) of the observations $\y_{1:\T}$, which is a common estimator used by practitioners to accommodate non-stationary data.   We interpret $\T_{\alpha}$ as the ``scaled count" of the number of observations in $\y_{1:\T}$, scaled by the weighting parameter $\alpha$.  

With prediction as our primary goal, the posterior predictive distribution of future observation $y^\star$ under our PWD approach is
\begin{eqnarray} \label{ppred-norm}
y^\star \, | \, \y_{1:\T}, \alpha \quad \sim \quad t_{\T_{\alpha} -1} \left( \hat{y}_{\alpha,\T} \, , \, \dfrac{\T_\alpha + 1}{\T_\alpha} S_{\alpha,\T} \right) \label{norm-postpred}
\end{eqnarray}
where 
\begin{eqnarray} \label{PWD-var}
S_{\alpha,\T} = \frac{\T_\alpha}{\T_\alpha-1} \left( \widehat{y_{\alpha,\T}^2} - \hat{y}_{\alpha,\T}^2 \right).
\end{eqnarray} 
The posterior distributions (\ref{norm-unknownvar-mean})-(\ref{norm-postpred}) reduce to the standard posterior distributions for a stationary model when $\alpha = 1$ whereas when $\alpha < 1$, data far in the past will be less relevant to the terminal time point and prediction of future observations.  

The posterior predictive distribution (\ref{norm-postpred}) has a very simple form that can be used to make predictions of future data $y^\star$ while avoiding the need to estimate the non-stationarity in the underlying parameters $\mu_t$ and $\sigma_t^2$ directly.  These results are conditioned on a known value of the weighting parameter $\alpha$ but in Section~\ref{selecting-alpha} we will discuss strategies for estimating $\alpha$ from the data.

%%%%%%%%%%%%%%%%%%%%%%%%%%%%%%%%%%%%%%%%%%%%%%%%%%
%\subsection{Comparison to Dynamic State Space Models} \label{SS-comparison}
\subsection{Related Time Series Approaches} \label{SS-comparison}

Our PWD approach for a normally distributed time series, $y_t \sim \mathcal{N} (\mu_t, \sigma_t^2)$,  closely mimics the first order state space model of \cite{west1998bayesian}, 
\begin{align*}
y_t &= \theta_t + \nu_t, \quad \nu_t \sim \mathcal{N} (0, V) \\
\theta_t &= \theta_{t-1} + \omega_t, \quad \omega_t \sim \mathcal{N} (0, W_t),
\end{align*}
with observation variance $V$ constant over time but state variance $W_t$ varying over time.  This state space model is estimated through recursive equations culminating in a normal posterior distribution for the terminal mean, $\theta_\T | \y_{1:\T} \sim \mathcal{N}(m_\T, C_\T)$ with
\begin{eqnarray*}
m_\T = m_{\T-1} + \frac{C_{\T-1} + W_\T}{C_{\T-1} + W_\T + V} (y_\T - m_{\T-1}) \quad {\rm and} \quad C_\T = \frac{C_{\T-1} + W_\T}{C_{\T-1} + W_\T+ V}
\end{eqnarray*}

\cite{west1998bayesian} also provide a {\it discounted} alternative to their model with {\it discount factor} $\delta \in [0,1]$ that  downweights more distant observations in the time series by inflating the posterior variance of $\theta_t$ at each time step $t$,   
\begin{eqnarray}  \label{const-disc}
m_\T = \dfrac{ \sum_{i=0}^{\T-1} y_{\T-i} \, \delta^i}{\sum_{i=0}^{\T-1} \, \delta^i} \quad \text{and} \quad C_\T = \dfrac{V}{\sum_{i=0}^{\T-1} \, \delta^i},
\end{eqnarray}
which are equivalent to our power-weighted densities (PWD) approach in (\ref{norm-unknownvar-mean})-(\ref{norm-unknownvar-var}).  However, this equivalence is specific to normally distributed data and does not hold for the more general PWD approach in (\ref{pp-eqn-general}).   

There is a similar connection between dynamic state space models and rolling windows approaches if the discount factor $\delta$ is allowed to vary over time in a lag-specific way with the following values, 
$$  \{ \delta_1, \delta_2, \cdots, \delta_{\T} \} = (\underbrace{0, 0, \cdots, 0}_{\T-p}, \underbrace{1, 1, \cdots, 1}_p).$$

%Harrison and West note that their discount factors can be made a function of time, replacing a constant $\delta$ with the sequence 
%\begin{equation}   \label{disc-general}
%\bdelta \stackrel{\Delta}{=} \{ \delta_1, \delta_2, \cdots, \delta_{\T} \}.
%\end{equation}
%Assuming a vague prior upon the mean, a more general version of Equation (\ref{const-disc}) where $\bdelta$ can vary as a function of time is
%\begin{equation}  \label{vary-disc}
%m_{\T} = \dfrac{\sum_{i=0}^{\T-1} \left[ \prod_{j=1}^{i} \delta_{\T-j} \right] y_{\T-i}}{ \sum_{i=0}^{\T-1} \prod_{j=1}^{i} \delta_{\T-j}}, \ \  C_{\T} = \dfrac{V}{\sum_{i=0}^{\T-1} \prod_{j=1}^{i} \delta_{\T-j}}.
%\end{equation}
%We can see that a length $p$ window given data up to time point $\T$ is equivalent to discount factors 
%$$  \{ \delta_1, \delta_2, \cdots, \delta_{\T} \} = (\underbrace{0, 0, \cdots, 0}_{\T-p}, \underbrace{1, 1, \cdots, 1}_p).$$
%If we further assume the discount factors are indexed by {\it lag} and not by {\it time}, this state space formulation is equivalent to a length $p$ rolling window as we move forward through time.  

This representation highlights two issues with rolling windows.  It is difficult to interpret rolling windows as a data generating process, since the normal model with a rolling window of length $p$ implies a posterior distribution for $\theta_t$ with infinite variance at all time points $t \in \{ 1, 2, \ldots, \T-p \}$.   It is also not clear how to estimate the optimal length $p$ of the rolling window.   

%These two issues are not shared by our exponentially weighted PWD approach.

We will see superior predictive performance of our PWD approach over discounted state space models and rolling windows in our stock market analysis in Section \ref{stock-mkt-ind}.   That said, we can still borrow insight from the discounted state space model of \cite{west1998bayesian} in terms of their estimation of the discount factor $\delta$.   In particular, they select $\delta$ which maximizes the one-step-ahead predictive likelihood of the data, and in Section~\ref{selecting-alpha}, we will employ a similar strategy for the estimation of our weight parameter $\alpha$.

%%%%%%%%%%%%%%%%%%%%%%%%%%%%%%%%%%%%%%%%%%%%%%%%%%
%\subsection{Comparison to Exponentially Weighted Moving Average Models} \label{ewma-comp}
%%%%%%%%%%%%%%%%%%%%%%%%%%%%%%%%%%%%%%%%%%%%%%%%%%

Our PWD approach for a normally distributed time series also bears similarity to the exponentially weighted moving average model (EWMA), also known as an autoregressive integrated moving average process, ARIMA (0,1,1), in which the first differences of the data are modeled as 
\begin{align}\label{eq:MAmod} 
y_t - y_{t-1} = \epsilon_t + \rho \, \epsilon_{t-1} \quad \text{ where } \quad \epsilon_t \sim \mathcal{N}(0,\sigma^2) \quad \text{ and } \quad \rho \in (-1,1)
\end{align}
Recursively applying equation \ref{eq:MAmod} and letting $\rho \equiv -\alpha$ gives 
\begin{align}  \label{eq:MAmod3}
y_t = \epsilon_t + (1-\alpha) \sum_{i=0}^{t-2} \alpha^i y_{t-i-1},
\end{align}
which has a similar mean for $y_t$ as the provided by our PWD approach in equation \ref{norm-mu-eqn}.   The $\alpha$ parameter is estimated via (in-sample) maximum likelihood estimation in the usual EWMA procedure, whereas in  Section \ref{selecting-alpha}, we propose estimating $\alpha$ by maximizing the {\it one-step-ahead predictive likelihood} of the data.   We will show substantial gains in terms of accuracy and computational cost of our PWD approach compared to EWMA in synthetic settings in Section \ref{const-mean}.   In addition, our PWD approach generalizes more naturally to hierarchical linear regression (Section \ref{hier-lin}) which is needed for our financial application as well as allowing for other decay specifications (such as rolling windows and linearly decaying weights).  

%In a simulation study in , this difference leads to a $~$20\% smaller average empirical root mean squared prediction error.  and its empirical average computing time was 5.4 times faster, despite EWMA's having been estimated using the {\tt arima} function in {\tt R's stats} package, which is optimized for speed.  
%While the traditional EWMA model would perform poorly in the regression-based simulation and empirical analyses we consider in Section \ref{hier-reg} and Section \ref{stock-mkt-ind} because it does not accommodate covariates, we include a close analog to the EWMA instead, the local level dynamic linear regression model, and show that our predictive performance is favorable.  Finally, our PWD approach generalized naturally to other decay specifications besides the exponential specification we focus upon for the majority of this paper (i.e., rolling window weights, ), as we alluded to in Section \ref{SS-comparison}.

%%%%%%%%%%%%%%%%%%%%%%%%%%%%%%%%%%%%%%%%%%%%%%%%%%
%\subsection{Other Related Literature} \label{other-comp}
%%%%%%%%%%%%%%%%%%%%%%%%%%%%%%%%%%%%%%%%%%%%%%%%%%

%Outside of the similarities to state space and EWMA models noted in Sections \ref{SS-comparison} and \ref{ewma-comp}, respectively, there is other related literature which merits mention.  

\cite{smith1979generalization} and \cite{smith1981multiparameter} introduce a Power Steady Model (PSM) which produces posterior distributions similar to our PWD approach for a general class of likelihoods with exponentially decaying weights.  \cite{grunwald1993time} extend \cite{smith1981multiparameter}'s framework to data which is conditionally Dirichlet-distributed.   However, in this approach both the likelihood and the prior distribution are power-weighted, whereas our PWD approach only power-weights the likelihood term.  It is also not clear how to extend this PSM model to non-exponential decays or hierarchical models.  

\cite{chen1994non} create a state-space model for data with a Gamma likelihood that includes a parameter for discounting older data in an exponential manner.   \cite{shephard1994local} derives state-space models with normal or exponential likelihoods where a scale parameter evolves over time.  
%In the normal likelihood case, the resulting estimate is roughly equal to the inverse of an EWMA of the square of the observations.   
Both of these approaches are distribution-specific and the entire evolution of the state variable is estimated, whereas our  PWD approach is intended as a fast and simple alternative to full state-space estimation when the goal is out-of-sample prediction.

%\cite{chen1994non} focus on estimating the model parameters at each time point using the Gibbs sampling algorithm, whereas our PWD approach focuses only upon estimation of the model parameters at the terminal time point and does not require Gibbs sampling.

%%%%%%%%%%%%%%%%%%%%%%%%%%%%%%%%%%%%%%%%%%%%%%%%%%
\subsection{Estimation of Weight Parameter $\alpha$} \label{selecting-alpha}
%%%%%%%%%%%%%%%%%%%%%%%%%%%%%%%%%%%%%%%%%%%%%%%%%%

Our estimation method for the weighting parameter $\alpha$ of our power-weighted densities approach mimics a method proposed by \citet{west1998bayesian} (p. 58) for their local level state-space model.  We select the value $\alpha^\star$ that maximizes the {\it one-step-ahead predictive likelihood}, 
\begin{eqnarray}
\qquad \alpha^\star \, = \, \underset{\alpha}{\text{argmax}}  \,\, p^\star(\alpha \, | \, \y) \, \stackrel{\Delta}{=} \, \underset{\alpha}{\text{argmax}}  \,\, p_0(\alpha) \, \prod_{t=2}^\T p(y_t \, | \, \y_{1:{t-1}},\alpha)  \label{alpha-argmax} 
\end{eqnarray}
with $p(y_t \, | \, \y_{1:{t-1}},\alpha)$ being the one-step-ahead predictive densities, 
\begin{eqnarray}
p(y_t \, | \, \y_{1:{t-1}},\alpha) = \int p(y_t \, | \, \btheta_t)  \, p_\alpha(\btheta_t \, | \, \y_{1:{t-1}})  \, p(\btheta_t)  \, d\btheta_t \label{predlik-int}
\end{eqnarray}
based on the power-weighted densities $p_\alpha(\btheta_t \, | \, \y_{1:{t-1}})$ from (\ref{pp-eqn}).    This procedure is consistent with our primary goal: prediction of the next time point.  The maximal value $\alpha^\star$ can be found with minor computational cost by a grid evaluation of $p^\star (\alpha \, | \, \y_{1:\T})$ over $\alpha \in [0,1]$, though it is often easier to maximize the logarithm of (\ref{alpha-argmax}) instead.   

Note that the predictive likelihood (\ref{alpha-argmax}) includes a prior distribution $p_0(\alpha)$ that can reflect any prior beliefs that a practitioner may have about the relative probability of particular values of $\alpha$.   In this paper, we will assume that all values of $\alpha$ are equally likely {\it a priori}.  

Our predictive likelihood approach is related to the model selection procedure of \citet{gelfand1994bayesian}.   Assuming all models in a set of candidate models are equally likely {\it a priori}, they propose selecting the model with the best $C$-fold cross-validated out-of-sample {\it forecasting accuracy}.   This strategy is also similar to the prequential approach of \cite{dawid1992prequential} where preference is given to estimators with the smallest predictive loss.

For the illustrative normal model estimated by (\ref{norm-unknownvar-mean})-(\ref{norm-postpred}), we select the $\alpha^\star$ that maximizes
\begin{eqnarray}  \nonumber
\log  p^\star(\alpha \, | \, \y) & = & \log p_0(\alpha) +  \sum_{t=2}^{\T-1} \log \Gamma \left(\frac{t_\alpha+ 1}{t_\alpha} \right)     - \frac{1}{2} \Big( \log(t_\alpha + 1) + \log S_{\alpha,t} \Big) \\    \label{ppred-lik}
 && - \Big( \dfrac{t_\alpha + 1}{2} \Big)  \log \Big( 1 +  \dfrac{ (y_{t+1} - \hat{y}_{\alpha,t+1})^2}{(t_\alpha + 1) S_{\alpha,t}} \Big), 
\end{eqnarray}
where $t_\alpha = \sum_{i=0}^{t-1} \alpha^i $, $\hat{y}_{\alpha,t} =  \sum_{i=0}^{t-1} y_{t-i} \, \alpha^i / t_\alpha$ and  
\begin{eqnarray*}
S_{\alpha,t} =  \frac{t_\alpha}{t_\alpha-1} \left( \widehat{y_{\alpha,t}^2} - \hat{y}_{\alpha,t+1}^2 \right) \quad {\rm with} \quad \widehat{y_{\alpha,t}^2} = \dfrac{\sum_{i=0}^{t-1} \alpha^i y_{t-i}^2}{t_\alpha} 
\end{eqnarray*}

 While the computation required for equation \ref{ppred-lik} may seem daunting, we show in \ref{suppA} that evaluation of this expression scales linearly with the length of the time series.   We will see in Section \ref{const-mean} that this linear time algorithm has computing times which are 5 to over 10 times faster than built-in {\tt R} functions exponential weighted moving average and state space implementations.  We will provide an {\tt R} package for our PWD approach so that practitioners may benefit from our fast implementation. 

One could also consider a fully Bayesian approach where we obtain posterior samples of $\alpha$ which would allow us to summarize the posterior variability in the weight parameter.   However, the estimated posterior distribution of $\alpha$ tends to favor $\alpha \rightarrow 0$ since small values of $\alpha$ correspond to individual parameters $\btheta_t$ for each observation $y_t$, since there is no penalty for over-parameterization when fitting the entire time series {\it in sample} through the posterior distribution.   For this reason, we prefer our one-step-ahead predictive likelihood approach (\ref{alpha-argmax}), since its out-of-sample nature inherently protects against over-parameterization.    If desired, we still can incorporate the variability in our weight parameter by instead sampling $\alpha$ from our one-step-ahead predictive likelihood $p^\star(\alpha \, | \, \y)$.  In \ref{suppA}, we present a simulation study that suggests a sampling approach for $\alpha$ does not lead to better predictive performance than using the point estimate $\alpha^\star$ from (\ref{alpha-argmax}).  

%%%%%%%%%%%%%%%%%%%%%%%%%%%%%%%%%%%%%%%%%%%%%%%%%%
\subsection{Power Weighted Densities for Hierarchical Linear Regression} \label{hier-lin}
%%%%%%%%%%%%%%%%%%%%%%%%%%%%%%%%%%%%%%%%%%%%%%%%%%
%Thus far, we have only outlined the specifics of our PWD approach for an illustrative example of a single time series of normally distributed data.   

In this section, we extend our power-weighted densities approach for a hierarchical linear regression model, which is necessary for our application to monthly industry portfolio returns in Section~\ref{stock-mkt-ind}.   For that analysis, we need to model multiple time series each with potentially differing degrees of non-stationarity, while sharing information hierarchically across the multiple stock portfolios.  

We consider the general setting of $J$ different time series with outcome $y_{j,t}$ and $p$ covariates $\X_{j,t}$ at each time point $t$ in group $j$.  We specify a separate regression model for each group $j$,
\begin{eqnarray}  
y_{j,t} = \X_{j,t} \, \bbeta_{j,t} + \epsilon_{j,t}, \quad \epsilon_{j,t} \sim \mathcal{N}(\0, \sigma_{j,t}^2) \label{hierarchicalregression}
\end{eqnarray}
with time varying coefficients $\bbeta_{j,t}$ and residual variances $\sigma_{j,t}^2$.   We share information across groups via a common prior distribution, 
\begin{eqnarray}  \label{priors-linreg}
\bbeta_{j,t} &\sim \mathcal{N}_{p} (\bbeta_0, \Sigma_0),
\end{eqnarray}
where $\Sigma_0$ is a diagonal matrix with diagonal entries $\btau^2$.   Note that by using a diagonal matrix $\Sigma_0$, we are assuming {\it a priori} independence of the components of $\bbeta_{j,t}$, but this still allows for {\it a posteriori} dependence.   We use non-informative prior distributions $p(\beta_{0,k} , \tau_k^2) \propto (\tau_k^2)^{-1/2}$ for our global parameters and $p (\sigma_{t,j}^2) \propto (\sigma_{t,j}^2)^{-1}$ for the residual variances.  

We can implement this hierarchical linear regression model using the Gibbs sampler (\cite{GemGem84}).  Denoting $\btheta_{-a}$ as all parameters excluding $a$, the conditional distributions of the global parameters for each covariate  $k=1,2,\ldots,p$ are
\begin{eqnarray}  \label{global-hyps}
\beta_{0,k}  | \, \btheta_{-\beta_{0,k}}, \y &\sim& \mathcal{N} \left( \frac{\sum_{j=1}^J \beta_{j,k}}{J} \, , \, \frac{\tau_k^2}{J}  \right), \nonumber \\ 
\tau_k^2 | \, \btheta_{-\tau_k^2}, \y &\sim& \text{InvGamma}\left(    \frac{J}{2} \, , \, \frac{1}{2} \sum_{j=1}^J (\beta_{j,k} - \beta_{0,k})^2    \right)  
\end{eqnarray}
 
If our hierarchical regression model was assumed to be {\it stationary} (i.e. $\bbeta_{j,t} = \bbeta_{j}$ and $\sigma_{j,t}^2 = \sigma_{j}^2$), we would have the following conditional distributions for the group-specific parameters, 
\begin{eqnarray}
\bbeta_{j} | \, \btheta_{-\bbeta_{j}}, \y & \sim & \mathcal{N}_{p} (\hat{\bbeta}_{j} \, , \, \hat{V}_{j}) \nonumber \\
\sigma_{j}^2 | \, \btheta_{-\sigma_{j}^2}, \y & \sim & \text{InvGamma} \left( \dfrac{\T}{2} \, , \,  \frac{1}{2} \sum_{i=1}^{\T} (y_{j,i} - \X_{j,i} \bbeta_{j})^2 \right) \label{hierreg-group-stationary}
\end{eqnarray}
where 
\begin{eqnarray*}
\hat{\bbeta}_{j} & = & \hat{V}_{j}  \left(({\sigma_{j}^2})^{-1} \X_{j,1:\T}^\prime \y_{j,1:\T} + \Sigma_0^{-1} \bbeta_0 \right)   \quad {\rm and} \\
\hat{V}_{j} & = & \left(({\sigma_{j}^2})^{-1} \X_{j,1:\T}^\prime \X_{j,1:\T})^{-1} + \Sigma_0^{-1} \right)^{-1} \, . 
\end{eqnarray*}

However, in our financial application (and for many other time series), the assumption of stationary in the group-specific parameters is not realistic.   Rather, we can use our exponentially-decreasing PWD approach (\ref{pp-eqn}) to address potential non-stationarity in our model parameters, 
\begin{equation} \label{pp-eqn-hierLR}
p_\alpha(\btheta_{j,\T} \, | \, \y_{j,1:\T}) \, \propto \, p_0(\btheta_\T) \prod_{i=0}^{\T-1} p(y_{j,\T-i} | \btheta_{j,\T})^{\alpha_j^i} \quad \ \alpha_j \in [0,1]
\end{equation}
where by using different weight parameters $\alpha_j$ we allow for differing degrees of non-stationarity in each time series $j$.   Under this PWD approach,  the conditional distributions of the {\it time-varying} group-specific parameters at terminal time point $\T$ are
\begin{eqnarray*}
 \bbeta_{j,\T} | \, \btheta_{- \bbeta_{j,\T}}, \y & \sim & \mathcal{N}_{p} (\hat{\bbeta}_{\alpha,j} \, , \, \hat{V}_{\alpha,j}) \nonumber \\
\sigma_{j,\T}^2 | \, \btheta_{-\sigma_{j,\T}^2}, \y & \sim & \text{InvGamma} \left( \dfrac{\T_{\alpha_j}}{2} \, , \,  \frac{1}{2} \sum_{i=0}^{\T-1} \alpha_j^i (y_{j,\T-i} - \X_{j,T-i} \bbeta_{j,\T})^2 \right) 
\end{eqnarray*}
where 
\begin{eqnarray*}
\hat{\bbeta}_{\alpha,j} & = & \hat{V}_{\alpha,j}  \left(({\sigma_{j,\T}^2})^{-1} \X_{j,1:\T}^\prime  \, \A_{j,\T} \,  \y_{j,1:\T} + \Sigma_0^{-1} \bbeta_0 \right)   \quad {\rm and} \\
\hat{V}_{\alpha,j} & = & \left(({\sigma_{j,T}^2})^{-1} \X_{j,1:\T}^\prime \, \A_{j,\T} \, \X_{j,1:\T} + \Sigma_0^{-1} \right)^{-1}
\end{eqnarray*}
with weighting matrix $\A_{j,\T} \stackrel{\Delta}{=} diag(1, \alpha_j, \alpha_j^2, \cdots, \alpha_j^{\T-1})$ and $\T_{\alpha_j} = \sum_{i=0}^{\T-1} \alpha_j^i$.  

Comparing to the stationary model (\ref{hierreg-group-stationary}), our PWD approach acts through the weight matrix $\A_{j,\T}$ to downweight observations that are farther away from terminal time point $\T$.  The global parameters $\bbeta_0$ and $\btau^2$ can still be sampled using (\ref{global-hyps}).

The model implementation above is conditional upon knowing the weight parameters $\alpha_j$ for each group.  
Our usual estimation procedure for the weight parameters (Section~\ref{selecting-alpha}) would be to select the $\alpha_j$ which maximizes the one step ahead predictive likelihood for each group $j$: 
\begin{eqnarray}
\qquad \alpha^\star_j = \, \underset{\alpha_j}{\text{argmax}}  \,\, p_0(\alpha_j) \, \prod_{t=2}^{\T} p_{\alpha_j}(y_{j,t} \, | \, \y_{j,1:t-1})  \nonumber
\end{eqnarray}
This requires the evaluation of each one-step-ahead predictive density
\begin{eqnarray}
\phantom{xxxxxxx} p_{\alpha_j}(y_{j,t} \, | \, \y_{j,1:t-1}) = \int p(y_{j,t}\, | \, \btheta_{j,t})  \, p_{\alpha_j}(\btheta_{j,t}\, | \, \y_{j,1:t-1})  \, p_0(\btheta_{j,t})  \, d\btheta_{j,t}  \label{ppred-ref} 
\end{eqnarray}
at each time point $t$ by integrating over posterior samples of $\btheta_{j,t}$, which becomes computationally intensive if there are many groups $J$.   

For that reason, we prefer the following approximate approach based on {\it plug-in estimators} of $\btheta$ which is very fast and performs well in practice.   Specifically, we estimate each $\alpha_j$ as 
\begin{eqnarray}
\qquad \alpha^\star_j = \, \underset{\alpha_j}{\text{argmax}}  \,\, p_0(\alpha_j) \, \prod_{t=2}^{\T} p_{\alpha_j,approx}(y_{j,t} \, | \, \y_{j,1:t-1}, \widehat{\btheta})  \label{alphamax-approx}
\end{eqnarray}
where $p_{\alpha_j,approx}(y_{j,t} \, | \, \y_{j,1:t-1}, \widehat{\btheta})$ is the predictive likelihood of $y_{j,t}$ using plug-in estimators of the model parameters.  For the hierarchical linear regression model, this predictive likelihood is 
\begin{eqnarray}
y_{j,t} \sim  t_{t_\alpha -p-1} \left( \X_{j,t} \tilde{\beta}_{j,t} \, , \, \tilde{\sigma}_{j,t}^2 + \tilde{V}_{j,t}   \right) \label{ppred-approx}
\end{eqnarray}
where
\begin{eqnarray*}
\tilde{\bbeta}_{j,t} & = & \tilde{V}_{j,t}  \left(({\tilde{\sigma}_{j,t}^2})^{-1} \X_{j,1:(t-1)}^\prime  \, \A_{j,t-1} \,  \y_{j,1:(t-1)} + \tilde{\Sigma}_0^{-1} \tilde{\bbeta}_0 \right),   \\
\tilde{V}_{j,t} & = & \left(({\tilde{\sigma}_{j,t}^2})^{-1} \X_{j,1:(t-1)}^\prime \, \A_{j,t-1} \, \X_{j,1:(t-1)} + \tilde{\Sigma}_0^{-1} \right)^{-1},  \\
 \tilde{\sigma}_{j,t}^2 & = & \left(  \sum_{i=0}^{t-1} \alpha_j^i (y_{j,t-i} - \X_{j,t-i} \tilde{\beta}_{j,t})^2  \right) / (\T_{\alpha_j} - p), \\
\tilde{\bbeta}_0 & = & \sum_{j=1}^J \tilde{\bbeta}_{j,t}/J, \quad \text{and} \quad \tilde{\Sigma}_0 = \sum_{j=1}^{J} (\tilde{\bbeta}_{j,t} - \tilde{\bbeta}_0)^2 / (J-1).
\end{eqnarray*}
with $t_{\alpha_j} = \sum_{i=0}^{t-1} \alpha_j^i $ and weighting matrix $\A_{j,t-1} = diag(1, \alpha_j, \alpha_j^2, \cdots, \alpha_j^{t-1})$.  Since each of the above plug-in estimators is a function of $\alpha_j$, we must iterate between: 
\begin{enumerate}
\item Updating the plug-in estimators $\tilde{\beta}_{j,t}$,  $\tilde{\sigma}_{j,t}^2$, $ \tilde{V}_{j,t}$, $\tilde{\bbeta}_0$ and $\tilde{\Sigma}_0^{-1}$ based on the current estimate of $\alpha_j$.
\item Optimizing $\alpha_j$ in (\ref{alphamax-approx}) using the predictive likelihood (\ref{ppred-approx}) based on the updated values of the plug-in estimators.  
\end{enumerate}
In \ref{suppA}, we show that evaluation of these expressions scale linearly with the length of the time series.  We assume convergence is achieved when the change in any $\alpha_j$ falls below a pre-specified threshold (in practice, we set this to be .005).  This plug-in method performs quite well in practice and has a much lower computation cost than the evaluation of integral (\ref{ppred-ref}) when estimating many group-specific $\alpha_j$'s.

%%%%%%%%%%%%%%%%%%%%%%%%%%%%%%%%%%%%%%%%%%%%%%%%%%
\subsection{Bayesian Model Averaging with Power Weighted Densities} \label{bma}

When modeling time series data, there is often uncertainty over the correct model to use in addition to the issue of non-stationarity within a particular model.   For example, in our financial application we consider the \citet{fama1993common} three factor model, but other popular alternatives are using no factors \citep{welch2008comprehensive}, the CAPM model, and the four factor model of \cite{carhart1997persistence} which adds an fourth momentum ('MOM') factor.   More generally, we may want to allow for any of the $2^4 = 16$ combinations of these four factors in our model for industry portfolios in Section~\ref{stock-mkt-ind}, and incorporate uncertainty about our model choices into our predictions.  

Bayesian model averaging (`BMA') is a popular way of allowing for model uncertainty \citep{kass1995bayes}, where the posterior densities of model parameters $\btheta$, are weighted by the probability of each model, $M_k$ $(k = 1, \ldots, K)$,
\begin{eqnarray} \label{bma-gen}
P(\btheta | D_{1:\T}) &=& \sum_{k=1}^{K} P(\btheta | D_{1:\T}, M_k) P(M_k | D_{1:\T}). 
\end{eqnarray}
with the weights proportionate to by the marginal likelihood of the data under each alternative model, 
\begin{eqnarray}  \label{bma-marglik}
P (M_k | D_{1:\T}) = \dfrac{P(D_{1:\T} | M_k)}{ \sum_{l=1}^{K} P(D_{1:\T} | M_l)}. 
\end{eqnarray}
with $D_{1:\T}$ denoting the data available up to and including time point $\T$.  

We adopt a predictive likelihood-based analog to BMA to allow for model uncertainty within our PWD approach.  Similar to how our PWD approach selects the value of $\alpha$ which maximizes the marginal one-step-ahead predictive likelihood of the observed data, our PWD-BMA approach weighs the posterior density of parameters $\btheta$ under each alternative models by their respective marginal one-step-ahead predictive likelihoods.  In other words, instead of Equation (\ref{bma-marglik}), we use 
\begin{eqnarray}
P_\alpha (M_k | D_{1:\T}) &=& \dfrac{ \prod_{t=2}^{\T} P(D_t | D_{1:t-1}, \alpha_k^\star, M_k) }{ \sum_{l=1}^K \prod_{t=2}^{\T} P(D_t | D_{1:t-1}, \alpha_l^\star, M_l) }, \label{bma-eq2}
\end{eqnarray} 
where $\alpha_k^\star$ maximizes the one-step-ahead marginal predictive likelihood of the data under model $M_k$:
\begin{eqnarray}
\alpha_k^\star = \text{argmax}_{\alpha} \prod_{t=2}^{\T} P(D_t | D_{1:t-1}, M_k).
\end{eqnarray}

%Weighing each of our $K$ models appropriately, we may easily compute the desired composite posterior density for $\bbeta_{\T}$ for each group $j$.  With this composite posterior density for $\bbeta_{j,\T}$ in hand for each group $j$, we may then predict what the return for that group will be at the next time point. 

BMA-based estimators have many favorable qualities \citep{raftery2003discussion} and tend to perform well in terms of  out-of-sample performance \citep{madigan1994model,hoeting2002bayesian}. In the finance literature, \cite{avramov2002stock} shows that BMA improves predictive regression forecast errors. \cite{rapach2009out} accommodates model uncertainty in a financial setting based upon \cite{stock2004combination}, by combining models using weights which are a function of their previous forecasting ability but with a discount factor which assigns greater weight to more recent forecasting accuracy.  \cite{aiolfi2006persistence} also address model uncertainty in a predictive regression setting, but \cite{rapach2009out} showed that their performance may be uneven when used to predict monthly equity returns.  \cite{dangl2012predictive} applied BMA to a state-space linear regression model and outperformed alternatives which do not allow for time-variation in the regression coefficients in a financial prediction setting.  

In summary, our PWD approach to model uncertainty is a variation of BMA where we, as in \cite{avramov2002stock}, weight each model by its predictive fitness, emphasizing more recent predictions more than older predictions which performed well in \cite{rapach2009out}.  We implement our approach in our financial application to industry stock portfolios in Section \ref{stock-mkt-ind}, which leads to both favorable performance and several implications for the importance of the model factors over time.

%%%%%%%%%%%%%%%%%%%%%%%%%%%%%%%%%%%%%%%%%%%%%%%%%%
\section{Simulation Evaluation of our PWD approach} \label{sim-eval}
%%%%%%%%%%%%%%%%%%%%%%%%%%%%%%%%%%%%%%%%%%%%%%%%%%

We use several synthetic data settings to evaluate the predictive and computational performance of our PWD approach relative to other methods.    We first consider a ``null" setting where the data are normally distributed with an underlying scalar mean that is {\it stationary} over time.   We then consider a {\it non-stationary} hierarchical regression setting that emulates the characteristics of our financial application in Section \ref{stock-mkt-ind}.  We also compare several variants of our PWD approach in simple non-stationary data settings in \ref{suppA}.

%%%%%%%%%%%%%%%%%%%%%%%%%%%%%%%%%%%%%%%%%%%%%%%%%%
\subsection{Stationary Normal Mean Setting}  \label{const-mean}
%%%%%%%%%%%%%%%%%%%%%%%%%%%%%%%%%%%%%%%%%%%%%%%%%%

While methods which allow for parameter evolution are expected to perform better when there is actual non-stationarity in those parameters, it is also important to evaluate performance of those methods when the underlying parameters are, in fact, stationary.   In this stationary case, non-stationary methods may lose predictive accuracy and have a higher computational cost.

We generate synthetic data for a univariate time series of length $\T=500$, where the true underlying mean of the time series is constant over time: 
\begin{eqnarray}
y_t = \beta + \epsilon_t \quad \text{where} \quad \epsilon_t \sim \mathcal{N}(0,\sigma^2).
\end{eqnarray}
We set the true mean $\beta = 2$ and variance $\sigma^2=1$.  We generate 4,000 datasets under this setting and use the first $\T-1$ time points of each dataset (holding out the terminal observation $y_\T$) to train the following models:
\begin{enumerate}
\item {\tt Stationary}: Assume mean is stationary and predict $y_\T$ with the simple average of the first $\T-1$ time points of each dataset.
\item {\tt PWD}: Predict $y_\T$ with the mean of the posterior predictive distribution from Equation \ref{ppred-norm} using   $\alpha^\star$ that maximizes Equation \ref{ppred-lik}.
\item {\tt EWMA}: Use {\tt R}'s ${\tt ARIMA}$ function within the {\tt stats} package to fit an ARIMA (0,1,1) model.  The prediction of $y_\T$ is an {\it exponentially weighted moving average} of the first $\T-1$ time points.
\item {\tt State-Space}:  Use {\tt R}'s ${\tt StructTS}$ function within the {\tt stats} package to fit a local level state space model via maximum likelihood. The prediction of $y_\T$ is the mean of the one-step-ahead predictive distribution.  
\end{enumerate}

In Table~\ref{stationarysim}, we compare these four methods in terms of root mean square prediction error (RMSE) for the held-out terminal observation $y_\T$ , the standard error (SE) over datasets of the RMSE, and the mean computing time in milliseconds (Time (ms)).  

\begin{table}[ht]
\centering
\begin{tabular}{|l|cccc|}
  \hline
 & {\tt Stationary} & {\tt EWMA} & {\tt PWD} & {\tt State-Space} \\ 
  \hline
RMSE & .045 & .064 & .054 & .064 \\ 
  SE & .000 & .001 & .001 & .001 \\ 
  Time (ms) & .01 & 6.13 & 1.14 & 11.52 \\ 
   \hline
\end{tabular}
\caption{Comparison of Methods in Stationary Setting} 
\label{stationarysim}
\end{table}
Since the underlying mean is stationary in this setting, {\tt Stationary} should have an advantage and this is indeed the case, with the RMSE for  {\tt Stationary} approximately 20\% lower than {\tt PWD}.  However, {\tt PWD} has an RMSE approximately 20\% smaller than both {\tt State-Space} and {\tt EWMA}, which suggests that our PWD approach is not as easily misled compared to these other methods when the underlying data generating process is truly stationary.  The small SEs suggest that all of these RMSE differences are statistically significant.  

\begin{figure}[ht!]
\centering
\includegraphics[width=70mm]{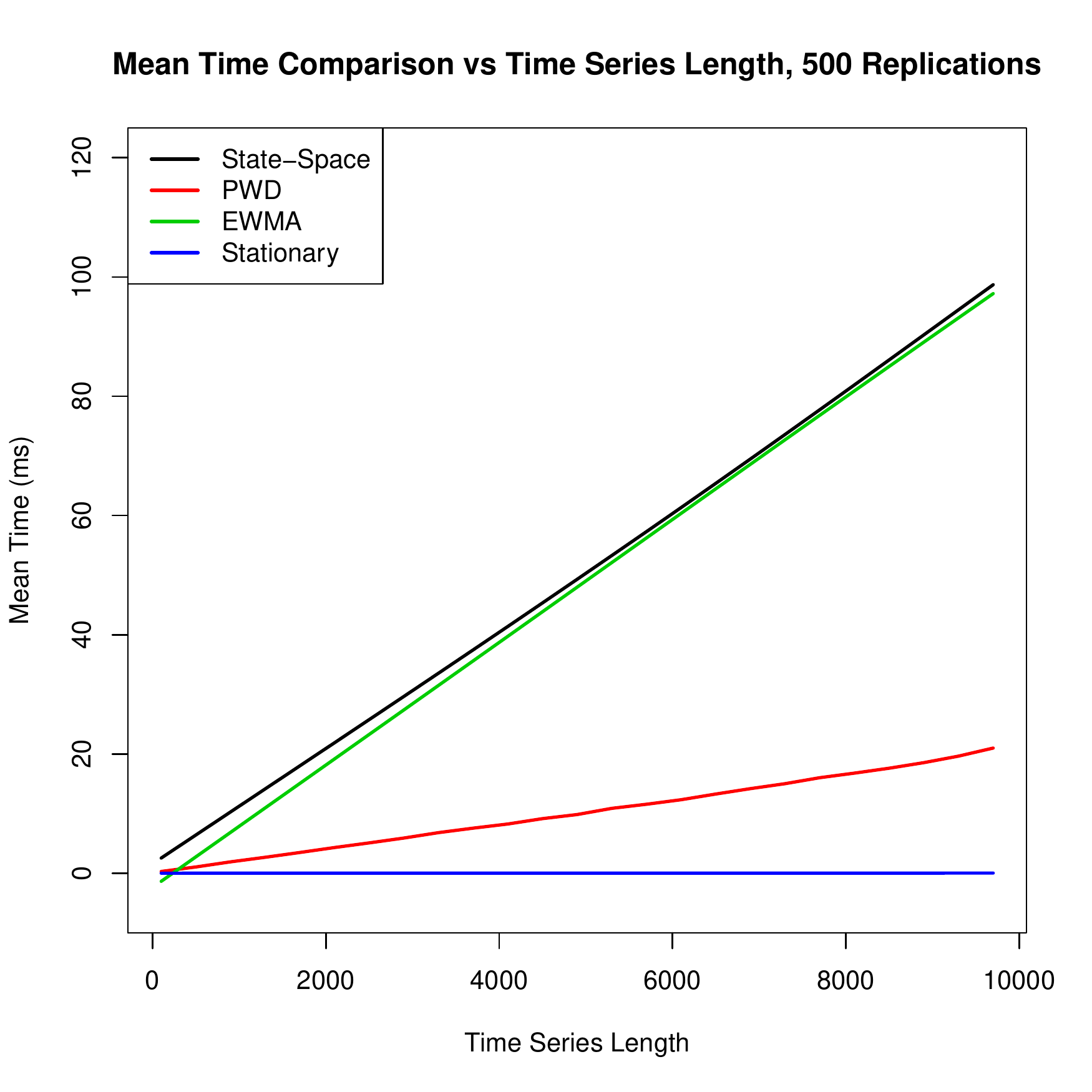}
\caption{\small{Mean Computing Time of 4 Models: {\tt Stationary}, {\tt PWD}, {\tt EWMA} and {\tt State-Space} as a function of time series length.}}
\label{time-comp}
\end{figure}

Moreover, {\tt PWD} was approximately 5 times faster than {\tt EWMA} and approximately 10 times faster than {\tt State-Space}.  This dramatic speedup is impressive given that the {\tt arima} and {\tt structTS} functions, as part of the {\tt stats} package within {\tt R}, have been optimized for speed.   In Figure  \ref{time-comp}, we further emphasize the reduced computational cost of our PWD approach by plotting the mean computing time in milliseconds (averaged over 2000 replications) for the four methods as a function of time series length.  All methods have computing times which scale linearly with time series length, but the slope associated with that linear scaling is much smaller for {\tt PWD} compared to {\tt EWMA} and {\tt State-Space}.

%%%%%%%%%%%%%%%%%%%%%%%%%%%%%%%%%%%%%%%%%%%%%%%%%%
\subsection{Non-Stationary Hierarchical Linear Regression Setting}  \label{hier-reg}
%%%%%%%%%%%%%%%%%%%%%%%%%%%%%%%%%%%%%%%%%%%%%%%%%%

We generate synthetic data in a regression setting that represents a simplified version of our financial application in Section  \ref{stock-mkt-ind}.  Specifically, each synthetic dataset consists of a set of $J$ portfolios
where the return on each portfolio $y_{j,t}$ is a linear function of the return of the overall market $m_t$, 
\begin{equation}  \label{ret-assumption}
y_{j,t} = \beta_{j,t} \, m_t + \epsilon_{j,t}  \qquad  \epsilon_{j,t}  \sim \mathcal{N}(0, \sigma^2) 
\end{equation}
with portfolio-specific sensitivities $\beta_{j,t}$ to the overall market that evolve over time $t$.  This synthetic data model is  analogous to the celebrated CAPM model \citep{fama1989business}.   The market factor is generated as $m_{j,t} \sim \mathcal{N}(\mu_m, \sigma_m^2)$ where we set $\mu_m = .047$ and $\sigma_m^2 = .045^2$ based on historical monthly stock market data from \cite{shiller2014}.  The sensitivity of stocks to the market is {\it non-stationary} in that $\beta_{j,t}$ is centered upon its value from the prior period, $\beta_{j,t-1}$, plus a disturbance term, 
\begin{equation}  \label{beta-assumption}
\beta_{j,t} = \beta_{j,t-1} + \eta_{j,t}.
\end{equation}
The evolution of $\beta_{j,t}$ is also {\it group mean reverting} in that the disturbance term $\eta_{j,t}$ pulls $\beta_{j,t}$ towards the group average of the prior period: 
\begin{equation}  \label{beta-assumption2}
\eta_{j,t} = \phi_{j} ( \bar{\beta}_{j,t-1} - \beta_{j,t-1}) + \zeta_{j,t}, 
\end{equation}
where $\phi_{j}$ represents the magnitude of stock $j$'s mean reversion towards the overall group average, $\bar{\beta}_{t-1}$ is the group average $\beta$ at time point $t-1$, and $\zeta_{j,t}$ is white noise:
\begin{equation}  \label{extra-terms}
\phi_{j} \sim \text{Beta}(a,b), \ \ \bar{\beta}_{t-1} = \sum_{j=1}^J \beta_{j,t-1}/J, \ \ \text{and} \ \ \zeta_{j,t} \sim \mathcal{N}(0,\tau^2).
\end{equation}

We set $\sigma^2 = .04^2$, $\tau^2 = .08^2$, $a = 3$ and $b=97$ which leads to a strong correlation between portfolios and the overall market and meaningful evolution of $\beta_{j,t}$ over time, as well as mild mean reversion, shrinking the market sensitivity of portfolios towards the group average of the prior time point, consistent with the notion of ``beta decay" amongst finance practitioners.   

We examine two different data settings using this particular data generating process. {\tt Setting 1} consists of a large number of groups ($J=100$) that each contain a short time-series ($\T = 10$).  {\tt Setting 2} consists of a small number of groups ($J=10$) that each contain a long time-series ($\T = 100$).  {\tt Setting 2} is more similar to our application to industry portfolios in Section~\ref{stock-mkt-ind}, as that data contains a small number of relatively long time series.  

We generated 500 synthetic datasets under both settings.   For each approach that we consider, we train the model on the first $\T-1$ observations of each time series, $\y_{j,0:\T-1}$, as well as the market return over that same time period, $\m_{0:\T-1}$, and then predict the terminal observation, $y_{j,\T}$ using the return on the market from the final time point, $m_{\T}$.  Performance of each method is judged based on the RMSE of that prediction.

In these evaluations, we consider two variants of our PWD approach that differ in the modeling of the weighting parameters and group-specific means: 1. {\tt Hier-PWD} where we model all portfolios simultaneously using the hierarchical linear regression model outlined in Section~\ref{hier-lin}, and {\tt Sep-PWD} where we model each portfolio separately without any sharing between portfolios.  

We compare these two PWD variants to three alternative approaches:
 \begin{enumerate}
\item {\tt Stationary:} estimate the parameters in (\ref{ret-assumption}) using standard OLS regression applied separately to each portfolio time series, assuming that the coefficients are stationary over time, i.e. $\beta_{j,t} = \beta_{j}$
\item {\tt Stationary-Hier:} estimate the parameters in (\ref{ret-assumption}) simultaneously across portfolios using the hierarchical linear regression model (\ref{ret-assumption})-(\ref{beta-assumption2}), but still assume the coefficients are stationary over time, i.e. $\beta_{j,t} = \beta_{j}$
\item {\tt State-space-LR:} estimate the coefficients $\beta_{j,t}$ in (\ref{ret-assumption}) using a local level dynamic linear regression model estimated via maximum likelihood \cite[]{petris2009dynamic}.
\end{enumerate}

Table \ref{bigJ-rmse} compares performance of our PWD variants to the three alternative approaches in {\tt Setting 1} where we have a large number of groups ($J=100$) that each contain a short time-series ($\T = 10$).  In this setting, there is limited data available within each portfolio time series to estimate non-stationary parameters, and so the hierarchical methods should benefit from borrowing strength between portfolios.  

\begin{table}[ht]
\centering
\begin{tabular}{|l|ccccc|}
  \hline
 & Hier-PWD & Sep-PWD & State-Space-LR & Stationary & Stat-Hier \\ 
  \hline
  Mean(RMSE) & 19.00 & 22.00 & 28.11 & 21.43 & 19.04 \\ 
  SE(RMSE) & 0.18 & 0.32 & 0.82 & 0.30 & 0.18 \\ 
  t-test p-value &  & 0.000 & 0.000 & 0.000 & 0.878 \\ 
   \hline
\end{tabular}
\caption{Comparison of Methods in {\tt Setting 1}: Large Number of Short Time Series} 
\label{bigJ-rmse}
\end{table}

In Table \ref{bigJ-rmse}, we evaluate each approach using the average RMSE of the terminal time point prediction across the 500 datasets, as well as the standard error of that average RMSE\footnote{RMSE(Mean) and RMSE(SE) are re-scaled by a factor of $10^4$.} .  Observing that {\tt Hier-PWD} had the best average RMSE, we also provide the p-value from a two-sided t-test (assuming unequal variances) of the difference between the RMSE of {\tt Hier-PWD} and the RMSE of each method.

%\begin{table}[ht]
%\centering
%\begin{tabular}{|l|ccccc|}
%  \hline
% & Hier-PWD & Sep-PWD & State-Space-LR & Stationary & Stat-Hier \\ 
%  \hline
%  Mean(RMSE) & 18.06 & 21.07 & 34.21 & 20.97 & 18.44 \\ 
%  SE(RMSE) & 0.23 & 0.38 & 1.50 & 0.37 & 0.25 \\ 
%  t-test p-value &  & 0.000 & 0.000 & 0.000 & 0.113 \\ 
%   \hline
%\end{tabular}
%\caption{RMSE: Large Group Count, Short Time Series 
%500 Replications versus Hier-PWD} 
%\label{bigJ-rmse}
%\end{table}

We see in Table \ref{bigJ-rmse} that {\tt Hier-PWD} performed significantly better (at the 1\% level) than all other methods except for {\tt Stationary-Hier}.  The fact that {\tt Stationary-Hier} was the only method competitive with {\tt Hier-PWD} suggests a benefit from sharing information across groups but perhaps not enough data within each group to benefit from allowing non-stationarity.   We note the particularly poor performance of {\tt State-Space-LR} in this data setting where we have a large number of short time series.  

Table \ref{smallJ-rmse} compares performance of our PWD variants to the three alternatives in {\tt Setting 2} where we have a small number of groups ($J=10$) that each contain a long time-series ($\T = 100$), which more closely emulates our financial application in Section~\ref{stock-mkt-ind} where we have long time series for a relatively small number of portfolios.  

\begin{table}[ht]
\centering
\begin{tabular}{|l|ccccc|}
  \hline
 & Hier-PWD & Sep-PWD & State-Space-LR & Stationary & Stat-Hier \\ 
  \hline
Mean(RMSE) & 18.52 & 19.14 & 18.83 & 20.67 & 20.52 \\ 
  SE(RMSE) & 0.27 & 0.29 & 0.27 & 0.37 & 0.37 \\ 
  t-test p-value & & 0.114 & 0.412 & 0.000 & 0.000 \\ 
   \hline
\end{tabular}
\caption{RMSE: Small Group Count, Long Time Series; 
500 Datasets} 
\label{smallJ-rmse}
\end{table}

%
%\begin{table}[ht]
%\centering
%\begin{tabular}{|l|ccccc|}
%  \hline
% & Hier-PWD & Sep-PWD & State-Space-LR & Stationary & Stat-Hier \\ 
%  \hline
%Mean(RMSE) & 19.23 & 19.15 & 18.96 & 20.95 & 20.86 \\ 
%  SE(RMSE) & 0.68 & 0.61 & 0.57 & 0.73 & 0.74 \\ 
%  t-test p-value & & 0.225 & 0.125 & 0.149 & 0.195 \\ 
%   \hline
%\end{tabular}
%\caption{RMSE: Small Group Count, Long Time Series; 
%500 Datasets} 
%\label{smallJ-rmse}
%\end{table}

%We again compare our methods using average RMSE (and its standard error) across the 500 datasets, as well as the p-value from a two-sided Welch t-tests of the difference between the {\tt Hier-PWD} and each method.   

Comparing between our two PWD variants, we see that the pooling induced by {\tt Hier-PWD} did not lead to as much of a gain in predictive performance as seen in {\tt Setting 1}.  The situation of few groups with substantial amounts of data within each group limits the benefit of hierarchically sharing information between groups.   In this long time series setting where there is ample data for estimating the non-stationary evolution of the underlying $\beta_{j,t}$'s, we see that the stationary models {\tt Stationary} and {\tt Stat-Hier} performed poorly relative to the non-stationarity methods.   Among the non-stationary methods, {\tt Hier-PWD}, {\tt Sep-PWD} and {\tt State-Space-LR} did not have significant differences in their predictive accuracy.  

Our evaluation of both {\tt Setting 1} and {\tt Setting 2} suggests our power-weighted densities approach is robust to different data conditions, and is especially beneficial in situations where information sharing between groups is important, as in financial markets.  The {\tt State-Space-LR} approach was less robust: it performed competitively in {\tt Setting 2} but performed significantly worse in {\tt Setting 1} where less data was available in each time series.

We also observed dramatic benefits of our PWD approach in terms of computational cost in both {\tt Setting 1} and {\tt Setting 2}.   Comparing the variant of our PWD approach most similar to the state-space model, {\tt Sep-PWD}'s average computing time was 20-40 times faster than {\tt State-Space-LR}.  {\tt Sep-PWD} had an average computing time for all groups of 50 milliseconds in {\tt Setting 1} (compared with 2099 milliseconds for {\tt State-Space-LR}) and 340 milliseconds in {\tt Setting 2}  (compared with 7660 milliseconds for {\tt State-Space-LR}). Indeed, our {\tt Sep-PWD} variant may strike the best balance between computing speed and predictive accuracy for practitioners.

%%%%%%%%%%%%%%%%%%%%%%%%%%%%%%%%%%%%%%%%%%%%%%%%%%
\section{Application to Prediction of Industry Portfolios} \label{stock-mkt-ind}
%%%%%%%%%%%%%%%%%%%%%%%%%%%%%%%%%%%%%%%%%%%%%%%%%%

The ability to accurately estimate the sensitivity of portfolios to market factors is very important to financial practitioners since it enables firms to more accurately `hedge' or decrease risk through offsetting financial positions.  Dynamic hedging forms the basis for the pricing of financial derivatives, and the expected cost of the dynamic replication of a financial derivative (as well as the variability) drives the cost that a financial institution will charge to sell that derivative \citep{wilmott1995mathematics}, directly tied to the notion of basis risk \citep{figlewski1984hedging}.   In this section, we apply our power-weighted densities (PWD) approach for hierarchical linear regression (Section \ref{hier-reg}) to estimate the sensitivity of industry stock portfolios to market factors over time, and compare with several alternative methods.  

Our data consists of 49 stock portfolios formed based upon industry, available on Kenneth French's website (\href{http://mba.tuck.dartmouth.edu/pages/faculty/ken.french/data_library.html#Research}{Kenneth French}).   Of those 49 industries, we restricted our attention to the portfolios with the longest time series: there are 30 industry portfolios with monthly data starting December 1932 and running through December 2014.  Using monthly data is the general convention in the CAPM and factor model literature (e.g. \cite{fama1989business} and \cite{lewellen2006conditional}).    

In total, we have $J = 30$ stock portfolios and $\T = 985$ monthly time points per stock portfolio with no missing data over that period.   This data provides us with a representative cross section of market returns for many different asset classes and is a similar setting to the ``few groups of long time series" synthetic {\t Setting 2} of Section~\ref{hier-reg}.  

The celebrated work of \cite{fama1993common} predicted the return $y_{j,t}$ on a stock portfolio $j$ at a time $t$ with a linear three factor model,
\begin{align} 
\phantom{y} y_{j,t} = \alpha_{j,t} + \beta^m_{j,t} \, \cdot \, m_{t} + \beta^s_{j,t}  \, \cdot \,  s_{t} + \beta^v_{j,t}  \, \cdot \,  v_{t} + \epsilon_{j,t}  \qquad \epsilon_{j,t} \sim \mathcal{N}(0,\sigma_{j,t}^2) \label{famafrench2}
\end{align}
where $m_{t}$ is the excess return on the market (MKT), $s_{t}$ is the excess return of small capitalization stocks over large capitalization stocks (SMB), and $v_{t}$ is the excess return of value stocks over growth stocks (HML).  Compared to equation (\ref{3factor-intro}), we are now specifying normally-distributed errors and allowing for coefficients that are possibly time-varying (e.g. $\beta^m_{j,t}$ rather than $\beta^m_{j}$, etc).  In the usual matrix notation, (\ref{famafrench2}) is
\begin{eqnarray} \label{mkt-eqn}
y_{j,t}  =  \X_{t} \cdot \bbeta_{j,t} + \epsilon_{j,t} \quad {\rm where} \quad \epsilon_{j,t} \sim \mathcal{N}(0,\sigma_{j,t}^2) \end{eqnarray}
with $\X_{t} = [1 \  m_{t} \  s_{t} \  v_{t}]$ and $\bbeta_{j,t} = [\alpha_{j,t} \  \beta^m_{j,t} \  \beta^s_{j,t} \  \beta^v_{j,t}]$.

It is reasonable to believe that the $\bbeta$'s for individual portfolios will have some central tendency, which suggests that sharing information across portfolios may be useful.  We share information between our set of $J=30$ portfolios through a global prior distribution at each time point,
\begin{eqnarray} \label{mkt-eqn-prior}
\bbeta_{j,t} \sim \mathcal{N} (\bbeta_{0,t}, \Sigma_{0,t}) \qquad j = 1, \ldots, J
\end{eqnarray}
with $\bbeta_{0,t} = [\alpha_{0,t} \  \beta^m_{0,t} \  \beta^s_{0,t} \  \beta^v_{0,t}]$ and $\Sigma_{0,t}$ being a diagonal matrix with diagonal elements $(\tau_{\alpha,t}^2 \   \tau_{m,t}^2  \ \tau_{s,t}^2 \ \tau_{v,t}^2 )$.  We use non-informative priors $p(\bbeta_{0,t} , \Sigma_{0,t}) \propto (\tau_{\alpha,t}^2 \, \tau_{m,t}^2 \, \tau_{s,t}^2 \, \tau_{v,t}^2)^{-1/2}$ for the global parameters as well as $p (\sigma_{t,j}^2) \propto (\sigma_{t,j}^2)^{-1}$ for the residual variances.  

Even with a hierarchical structure on the parameters, this model is difficult to estimate unless we make a strong assumption of {\it stationarity} over time, i.e. $\bbeta_{j,t} = \bbeta_{j}$ for all $t = 1,\ldots,\T$.   However, stationarity is not a reasonable assumption in most financial applications, and the standard approach in the literature (e.g. \cite{fama1993common}) is to estimate time-varying coefficients using a rolling window.  

As an alternative to rolling windows, we will apply our power-weighted density (PWD) approach for hierarchical linear regression models (Section~\ref{hier-lin}) to this set of 30 industry portfolio time series.  Our PWD approach allows the regression coefficients $\bbeta_{j,t}$ to evolve over time for each portfolio $j$ but avoids estimating the entire parameter vector $\bbeta_{j,1:\T}$ when constructing the posterior distribution for the terminal coefficients $\bbeta_{j,\T}$ that are  used to predict the future return $y_{j,\T+1}$.   As outlined in Section~\ref{hier-lin}, we estimate portfolio-specific weighting parameters $\alpha_j$ so that influence of past observations can vary between different portfolios.  

We apply three variants of our PWD approach.  The first two variants, {\tt Hier-PWD} and {\tt Sep-PWD}, were employed in our synthetic data evaluation in Section \ref{hier-reg}.   We also consider a third variant, {\tt Sep-PWD-BMA}, where we combine Bayesian model averaging with our PWD approach as described in Section \ref{bma}.   This BMA variant explores 16 different linear models that are the combinations of inclusion/exclusion of the three Fama-French factors and the extra momentum factor of \cite{carhart1997persistence}.   We will evaluate the quality of these different models for each industry stock portfolio $j$ at each time point $t$.   The fast computational speed of our PWD approach greatly aids the practical implementation of {\tt Sep-PWD-BMA}. 

We will compare our three PWD variants to several alternative time series methods.  Three of these alternatives were also evaluated in Section \ref{sim-eval}: \\ {\tt Stationary}, {\tt Stationary-Hier} and {\tt State-space-LR}.  We will also evaluate a rolling window approach,  {\tt Window-5}, which estimates the coefficients $\bbeta_{j,t}$ in (\ref{mkt-eqn}) at each time point $t$ with a standard OLS regression using only the 5 years prior to time point $t$, same as in \cite{petkova2005value}.   Rolling windows are the standard approach to non-stationarity in the financial literature \cite[]{welch2008comprehensive}.  

We will evaluate the predictive performance of each method by the rolling cumulative evaluation of their forecast errors, as done in \cite{welch2008comprehensive}.  Specifically, for a particular model $M$ and a specific portfolio $j$ up to time point $t$, we calculate the squared prediction error between the actual return at time $t+1$ and predicted return given all information up to time $t$, 
\begin{eqnarray}   \label{spe-formula}
SPE(M)_{j,t+1} & = & \left(y_{j,t+1} - \hat{y}_{j,t+1} \right)^2  \nonumber \\
 & = & 
\left(y_{j,t+1} - \widehat{\alpha}_{j,t} - \widehat{\beta^m}_{j,t} \, m_{j,t+1} - \widehat{\beta^s}_{j,t} \, s_{j,t+1} -\widehat{\beta^v}_{j,t} \, v_{j,t+1}  \right)^2 \label{spe-eqn} 
\end{eqnarray}
where $(\widehat{\alpha}_{j,t},  \widehat{\beta^m}_{j,t}, \widehat{\beta^s}_{j,t}, \widehat{\beta^v}_{j,t})$ are estimated by model $M$ using all data up to time point $t$.   We aggregate the squared prediction errors across all $J=30$ stock portfolios to get the {\it cumulative sum of squared prediction errors} for a particular model $M$ up to any time point $t$, 
\begin{eqnarray*}
SSPE(M)_{1:t} = \sum_{i=1}^{t} \sum_{j=1}^J SPE(M)_{j,i}.
\end{eqnarray*}
For each model, we evaluate this cumulative sum of squared prediction errors at each monthly time point, starting in November 1937 when all competing methods are able to provide predictions, and ending in December 2014.   As in \cite{welch2008comprehensive}, we select the {\tt Stationary} model as a benchmark for our comparison since it represents the simplest approach to estimating the \cite{fama1993common} three-factor model.  Relative to the benchmark {\tt Stationary} model, we can calculate the difference in our cumulative sum of squared prediction errors up to time point $t$, 
\begin{eqnarray*}  \label{delta-ss}
\Delta SSPE(M, {\tt Stationary})_{1:t} = SSPE(M)_{1:t} - SSPE({\tt Stationary})_{1:t}
\end{eqnarray*}

\begin{figure}[ht!]
\centering
\includegraphics[width=100mm]{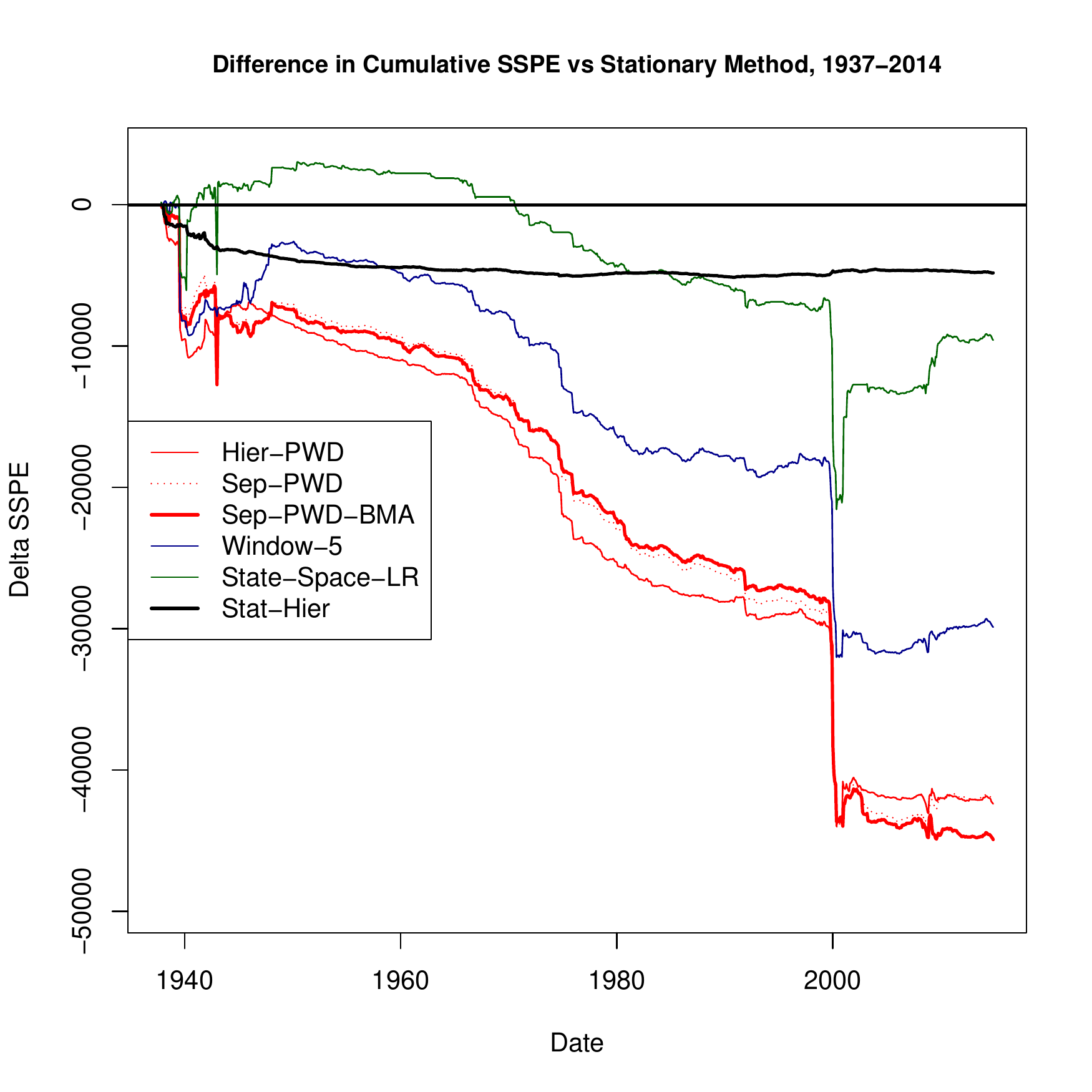}
\caption{\small{Rolling $\Delta SSPE$ relative to {\tt Stationary} model of six models: {\tt Hier-PWD}, {\tt Sep-PWD}, {\tt Sep-PWD-BMA}, {\tt Window-30}, {\tt State-space-LR}, and {\tt Stat-Hier}.}}
\label{sse-50yrs}
\end{figure}

In Figure \ref{sse-50yrs}, these differences in the cumulative sum of squared prediction errors (relative to {\tt Stationary}) are plotted over time for our three PWD variants and our alternative models.  

The most striking feature of Figure~\ref{sse-50yrs} is that the non-stationary methods ({\tt Window-5}, {\tt Sep-PWD}, {\tt Hier-PWD} and {\tt Sep-PWD-BMA}) show much better predictive performance than the baseline {\tt Stationary} model, with rolling cumulative prediction errors $\Delta SSPE$ that grow increasingly negative over time.   {\tt State-space-LR} model shows the worst predictive performance among the non-stationary methods.     {\tt Stationary-Hier} offers even less improvement over the stationary model, although we do see some gains predictive performance from the hierarchical version of the stationary model.

Among the non-stationary methods, the three variants of our power-weighted densities approach, {\tt Sep-PWD}, {\tt Hier-PWD} and {\tt Sep-PWD-BMA}, have the best predictive performance with increasingly lower cumulative prediction errors than the rolling window ({\tt Window-5}) and dynamic linear model ({\tt State-space-LR}) methods.  The outperformance of our PWD approach is not isolated to any one period of time, though the time period around 2000-01 saw a sharp jump in the gains for all non-stationary methods.

In Table \ref{industry-RMSE} we evaluate each model $M$ by its squared prediction error, $SSPE(M)$.  We provide the mean  $SSPE(M)$ averaged over time points and portfolios as well as its standard error across portfolios. Observing that {\tt PWD-BMA} had the smallest mean $SSPE(M)$, we also provide the p-value for a t-test on the difference between the {\tt PWD-BMA} mean $SSPE$ and the mean $SSPE$ of each other method.  

\begin{table}[ht]
\centering
\begin{tabular}{rrrrrrrrr}
\hline
Statistic & PWD-BMA & H-PWD & Sep-PWD & SS-LR & W-5 & Stat & H-Stat \\
\hline
Mean & 13392 & 13476 & 13481 & 14570 & 13893 & 14889 & 14729\\
Std. Error & 367 & 358 & 363 & 392 & 378 & 360 & 353 \\
p-value & & 0.619 & 0.233 & 0.001 & 0.000 & 0.000 & 0.000 \\
\hline
\end{tabular}
\caption{Industry Portfolio Performance Comparison: Squared Prediction Error Mean, Standard Error, and p-value of Difference in Mean versus PWD-BMA. For compactness, we denote Hier-PWD by H-PWD, State-space-LR by SS-LR, Window-5 by W-5, Stationary by Stat and Hier-Stat by H-Stat} 
\label{industry-RMSE}
\end{table}

Table \ref{industry-RMSE} implies that {\tt PWD-BMA}, {\tt Hier-PWD} and {\tt Sep-PWD} significantly improved upon the performance of {\tt State-space-LR}, {\tt Window-5}, {\tt Stationary}, {\tt Stat-Hier} and {\tt Stat-BMA}.  We see that {\tt PWD-BMA} outperformed all other methods, achieving the smallest mean squared prediction error as well as one of the smaller standard errors.  Financial practitioners value improvement in both the mean and the variance of squared prediction error because both reduce the amount of capital a financial practitioner would need to hold aside to maintain a hedge position over time.  As we will see shortly, it appears {\tt PWD-BMA} was able to adapt to secular cycles in the importance of the different market factors.  

%%%%%%%%%%%%%%%%%%%%%%%%%%%%%%%%%%%%%%%%%%%
\subsection{Evolution of $\alpha_j^\star$ and $\beta_j^m$ over time}

Our power-weighted densities approach provides some additional insight when we compare the estimated weight parameters $\alpha^\star_j$ for each of the 30 industry portfolios and their implication for the evolution of the sensitivities to changes in the overall stock market (``market beta").  

\begin{figure}[ht!]
\centering
\includegraphics[width=80mm]{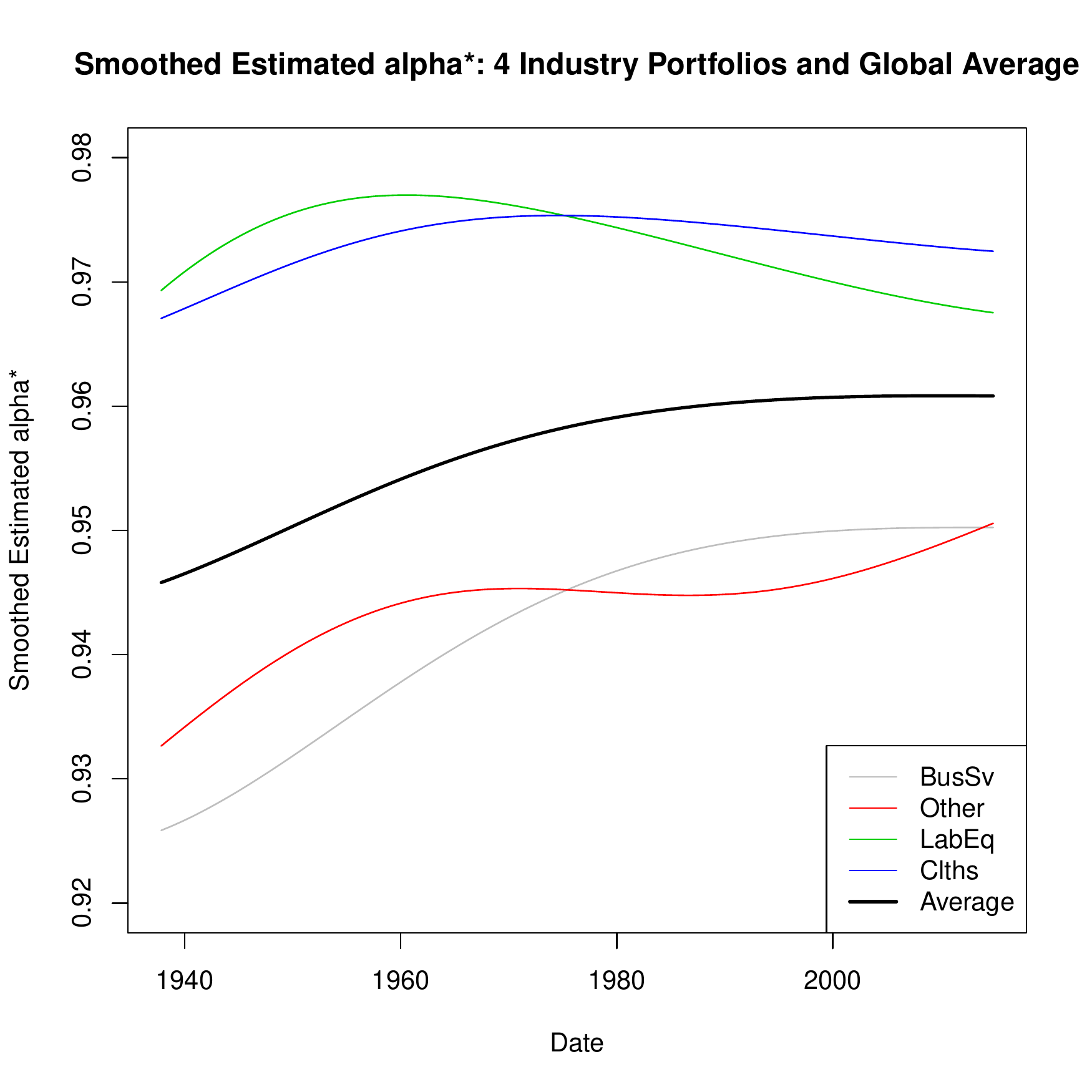}
\caption{\small{Smoothed estimated $\alpha^\star_j$ for highest and lowest empirical average estimated $\alpha_j^*$ over time. A local linear kernel bandwidth smoother over a dense grid of 600 grid points was used for the smoothing.
} }
\label{alpha-star-roll}
\end{figure}

In Figure \ref{alpha-star-roll}, we compare the estimated $\alpha^\star_j$  for the two industry portfolios with the lowest average $\alpha^\star_j$ to the two industry portfolios with the highest average $\alpha^\star_j$, as well as  the average $\alpha^\star_j$ across all portfolios.  Each $\alpha^\star_j$ is plotted as a smoothed trend over time, where the value of $\alpha^\star_j$ at time $t$ is estimated using data for that portfolio up to time $t$.  

The average $\alpha^\star$ across industries has been trending slightly upwards over time. Business services and other industries (``BusSv" and ``Other'') have the highest amount of non-stationarity (lowest $\alpha_j^\star$ values) whereas lab equipment and clothes (``Lab Eq" and ``Clths") have the lowest amount of non-stationarity (highest $\alpha_j^\star$ values).  It is unsurprising that the ``Other" industry has high non-stationarity since its risk profile and industry mix is most likely to change over time, while an industry like "Clothes" has a more stable risk profile over time.

\begin{figure}
\centering
\begin{minipage}{0.45\linewidth}
\centering
\includegraphics[width=65mm]{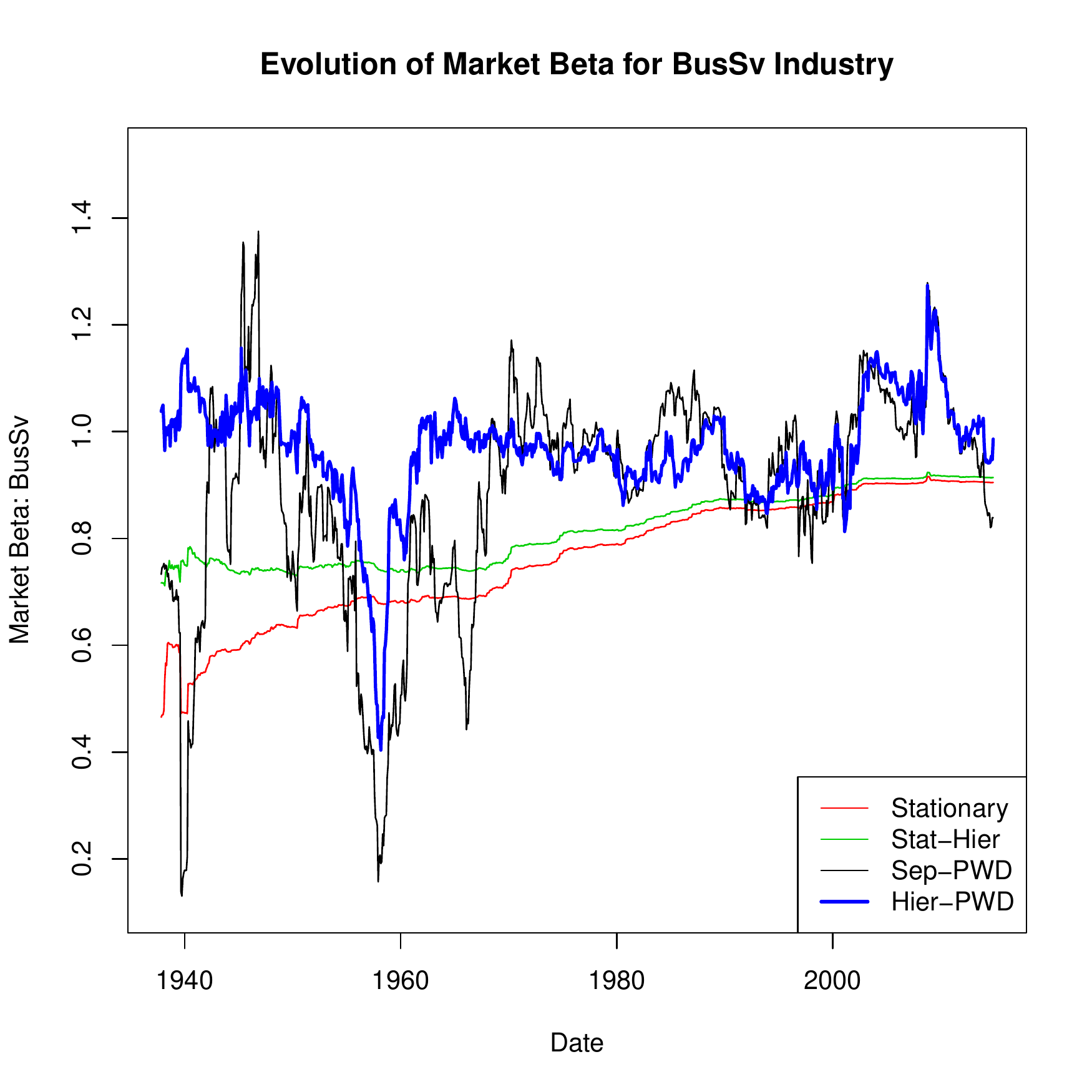}
\caption{\small{Estimated $\hat{\beta}^m_j$ for BusSv industry over time from 4 models} }
\label{BusSvRollBeta}
\end{minipage}\hfill
\begin{minipage}{0.45\linewidth}
\centering
\includegraphics[width=65mm]{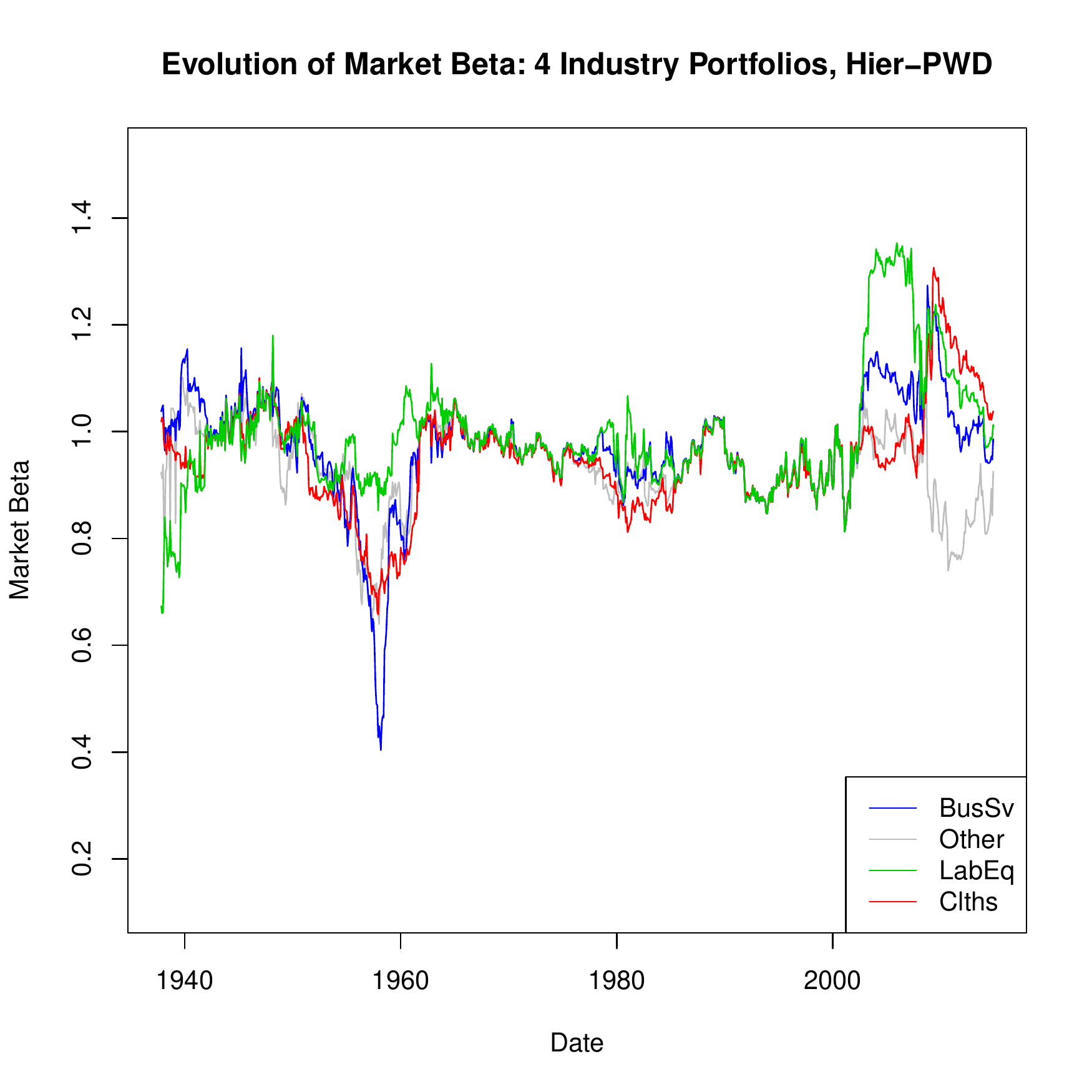}
\caption{\small{Estimated $\hat{\beta}^m$ from {\tt Hier-PWD} for four industries over time} }
\label{IndsRollBeta}
\end{minipage}
\end{figure}

Figure \ref{BusSvRollBeta} provides further examination of the role of our PWD weighting on $\beta^m$, the sensitivity of industry portfolio returns to the overall market over time.  Specifically, we plot the estimated of $\beta^m_j$ over time for the Business Services industry as estimated by the {\tt Stationary}, {\tt Stat-Hier}, {\tt Sep-PWD} and {\tt Hier-PWD} models.   Our PWD approaches suggest that $\beta^m_j$ for the Business Services industry is far less stable over time than implied by the {\tt Stationary} and {\tt Stat-Hier} models.   For example, after the burst of the technology stock market bubble in the early 2000's, our PWD approach inferred a sharp rise in $\beta^m_j$ which is indicative of heightened sensitivity of returns to overall market movements, while the stationary models made no such adjustment.   

Figure \ref{IndsRollBeta} compares the evolution of $\beta_j^m$ estimated by our {\tt Hier-PWD} approach for the four industries that represented the highest and lowest degrees of non-stationarity in Figure \ref{alpha-star-roll}.  Interestingly, there were two time periods in which $\beta_j^m$'s sharply diverged from 1.0: the period preceding 1960 and immediately following 2000.   These fluctuations would not be detectable by a stationary model that uses all historical data to estimate $\beta_j^m$. 

%%%%%%%%%%%%%%%%%%%%%%%%%%%%%%%%%%%%%%%%%%%
\subsection{The Evolution of Factor Weightings in Bayesian Model Averaging}

The Bayesian model averaging variant of our PWD approach provides additional insight into the importance of the different predictor factors over time and across industries.  As outlined in Section~\ref{bma}, our PWD-BMA calculates posterior model probabilities (equation~\ref{bma-eq2}) for the $16$ possible models that can be formed by the inclusion/exclusion of our four factors.   We calculate the {\it posterior probability of inclusion} for each factor as the sum of the posterior model probabilities over the subset of models that included that factor.   These inclusion probabilities are calculated for each portfolio $j$ and for each time point $t$ (using only data up to that time point).  

%As in \cite{avramov2002stock}, the basis for our analysis involves computing the rolling posterior probabilities of each explanatory factor for each stock portfolio at every time point predicted.  For stock portfolio $j$, we denote this by $W_j$, a $p$ x $\T$ matrix for which 
%$$W_j \equiv A^T P_j, $$
%where $A$ is a $2^p$ x $p$ matrix representing all forecasting models by zeros and ones which denote the exclusions and inclusions of factors, and $P_j$ is a $2^p$ x $\T$ matrix containing the posterior probabilities for stock portfolio $j$ of our $2^p$ regression models at each of the $\T$ time points that we provide predictions for.  
%This implies that if the intercept model was given all posterior probability for stock portfolio $j$, $W_j$ would be a matrix containing all zeroes.  If instead the three and four factor models equally shared all posterior probability mass over all time points predicted, the first three rows of $W_j$ would have all elements equal to 1.0, while the fourth row would have all elements equal to 0.5. 
%Denote the posterior probability of the $k$th explanatory factor for stock portfolio $j$ at each time $t$ by $w_j^k(t)$.  

\begin{figure}
\centering
\begin{minipage}{0.45\linewidth}
\centering
\includegraphics[width=65mm]{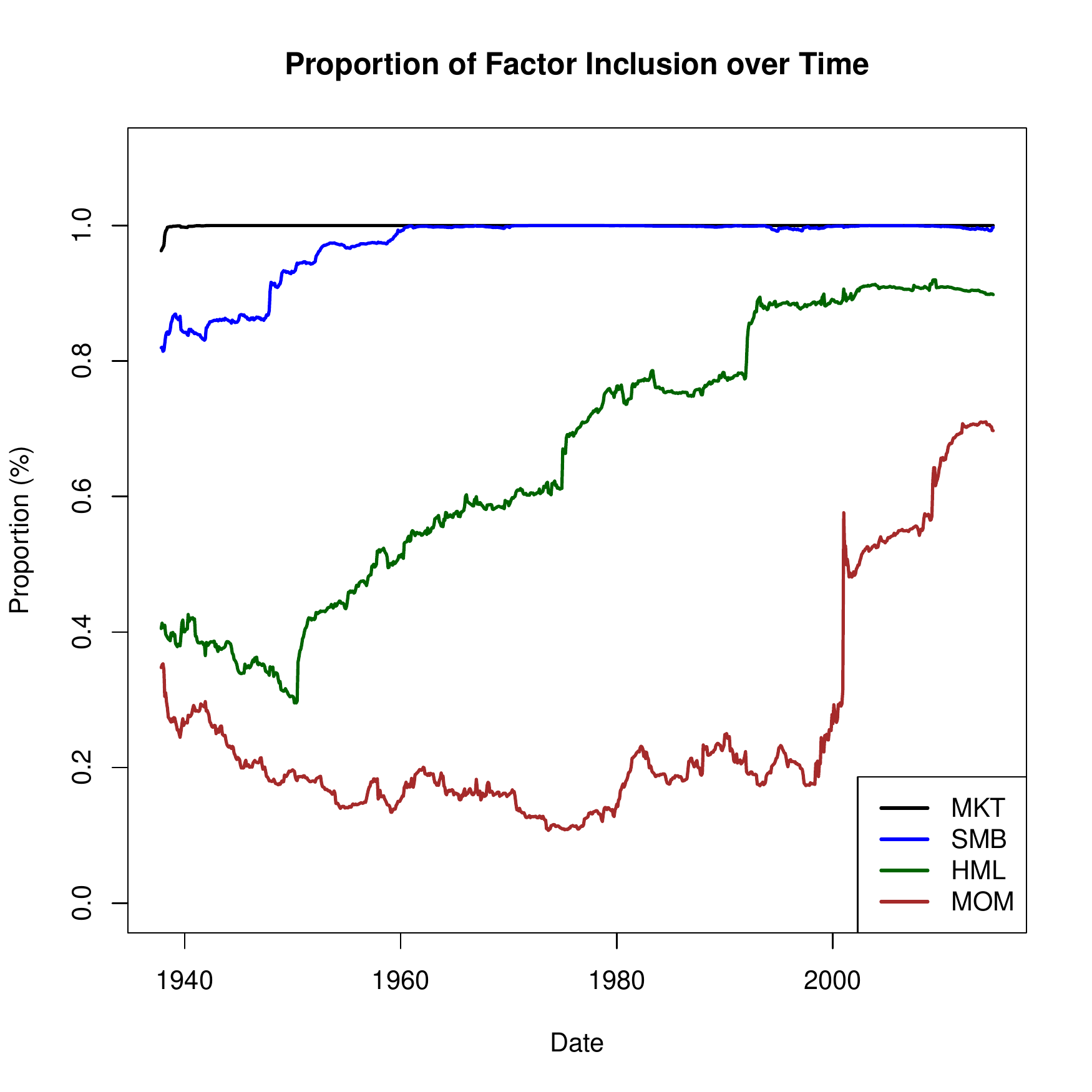}
\caption{\small{Inclusion probability for each factor, averaged over portfolios} }
\label{bma-time}
\end{minipage}\hfill
\begin{minipage}{0.45\linewidth}
\centering
\includegraphics[width=65mm]{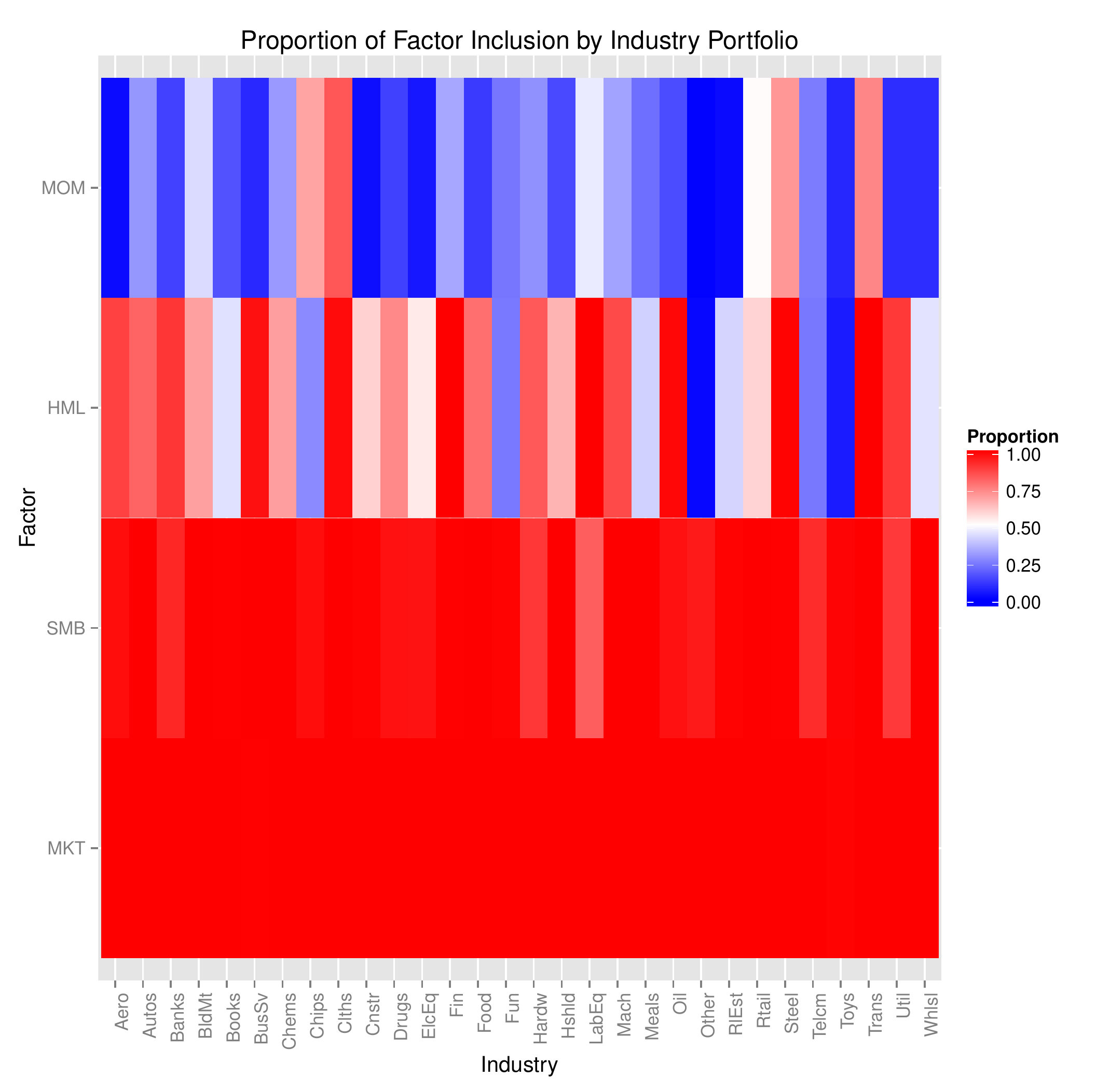}
\caption{\small{Inclusion probability for each factor by industry portfolio} }
\label{bma-port}
\end{minipage}
\end{figure}

In Figure \ref{bma-time} we plot the evolution of the inclusion probability, averaged over the thirty portfolios, for each of the four factors: MKT, SMB, HML and MOM.  In Figure \ref{bma-port}, we give the inclusion probability for each of the four factors averaged over time separately for each of the thirty portfolios.  

We see in Figure \ref{bma-time} that the MKT and SMB factors have inclusion probabilities near to 1.0 for almost the entire time series.  The HML and MOM factors initially have much lower inclusion probabilities for most of the time series, with the momentum factor being particularly interesting.  For almost 60 years, MOM's inclusion probability vacillated between 15\% and 25\% with an inclusion probability of 17\% in November 1997.   The MOM inclusion probability abruptly increased to 58\% by January 2001 and then further increased to 70\% by November 2014.   It is probably not coincidence that \cite{carhart1997persistence}, which first introduced the momentum factor, immediately preceded a sharp rise in the importance of MOM after 60 years of relative unimportance.   We observe a similar phenomenon with the HML factor, which had a step function-like increase from 78\% in December 1991 to 89\% in January 1993, the month before \cite{fama1993common} was published.  

Of the 16 regression models considered, the one with the highest posterior probability over the time period from November 1937 to May 1960 was a two factor model including only MKT and SMB factors.  From June 1960 to January 2001, the three factor model had the highest posterior probability.  Thereafter, the four factor model had the highest posterior probability.  
We see in Figure \ref{bma-port} the considerable heterogeneity across industries in the inclusion probabilities of the MOM and HML factors.  For example, the ``Other'', ``Aerospace" and ``Real Estate" industries have MOM inclusion probabilities of under 2\%, while the ``Steel", ``Transportation" and ``Clothes" industries have MOM inclusion probabilities above 70\%.  

These results could impact how financial practitioners may want to go about hedging their positions.  For example, our PWD approach does suggest that MOM factor is more important than it has ever been in explaining cross-sectional heterogeneity in returns across stock portfolios. 

%%%%%%%%%%%%%%%%%%%%%%%%%%%%%%%%%%%%%%%%%%%%%%%%%%
\section{Summary and Discussion}  \label{discussion}
%%%%%%%%%%%%%%%%%%%%%%%%%%%%%%%%%%%%%%%%%%%%%%%%%%

As an alternative to standard times series models, we have developed a power-weighted densities (PWD) approach where observations in the distant past are down-weighted in the likelihood function relative to more recent observations (\ref{pp-eqn-general}).   Our general approach provides an effective way to allow for non-stationarity in time series while still giving the practitioner control over the choice of the underlying data model, which could be useful in a wide variety of applications.   In this paper, we focused on a specific exponentially-decreasing weighting scheme (\ref{pp-eqn}) though other weighting schemes could be considered.  For example, the most popular way of allowing for non-stationarity in finance, rolling window estimation, is another special case of our PWD approach.

Our PWD approach is a simpler alternative for allowing non-stationarity compared to dynamic linear state space methods \citep{west1998bayesian} that explicitly model the evolution of an underlying state vector.  Our approach has the greatest benefit when the goal is forward-looking prediction, which is relevant in our application: prediction of future prices given the concurrent movement of market factors is often the primary goal in the financial markets.  With this emphasis on prediction, we have focused heavily on the posterior distribution for the parameters $\btheta_\T$ at the terminal time point $\T$, instead of inferring the entire evolution of an underlying state vector $\btheta_{1:T}$ as is done in state space models.  

Our simulation evaluation (Section \ref{sim-eval}) suggests that our PWD approach performs well in terms of both predictive accuracy and computational cost across different data settings and should be considered in situations where the practitioner suspects the underlying process generating the data evolves over time.  

In Section~\ref{hier-lin}, we developed the specific methodology for our PWD approach for a hierarchical linear regression model, which was needed for our application to industry portfolios in Section~\ref{stock-mkt-ind}.  In that application, our PWD approach showed superior predictive performance over models that assume stationary parameters, as well as alternative non-stationary methods such as dynamic linear models and rolling windows.  In Section \ref{bma}, we developed a PWD variant of Bayesian Model Averaging which yielded the best predictions in our application, and also gave interesting insights into the evolution in the importance of market factors over time. 

%%%%%%%%%%%%%%%%%%%%%%%%%%%%%%%%%%%%%%%%%%%%%%%%%%
\section{Acknowledgements}
Thanks to the Wharton Research Computing team for their HPC resources and to Joshua Magarick, Tengyuan Liang, Robert Stine and Nathan Stein for helpful discussions.

\begin{supplement} 
\sname{Supplement A}  \label{suppA}
\stitle{Discussion of ``Improving Market Factor Estimation with Power Weighted Densities"}  
%\sdatatype{.pdf}
\sdescription{We show the conjugacy for exponential families under our PWD approach and the Kullback-Leibler optimality of the general PWD setup.  We provide additional results for computational cost and simulations comparing additional PWD variants to competing models.  An adaptive PWD variant which switches between linear and exponentially decaying weights is also explored.}
%\slink[url]{http://www.e-publications.org/ims/support/dowload/imsart-ims.zip} \\
\end{supplement}

\bibliographystyle{imsart-nameyear}
\bibliography{biblio}

\begin{thebibliography}{37}
% BibTex style file: imsart-nameyear.bst, 2013-01-28
% Default style options (sort=1,type=nameyear).
% Used options (sort=1,type=nameyear).

\bibitem[\protect\citeauthoryear{Aiolfi and
  Timmermann}{2006}]{aiolfi2006persistence}
\begin{barticle}[author]
\bauthor{\bsnm{Aiolfi},~\bfnm{Marco}\binits{M.}} \AND
  \bauthor{\bsnm{Timmermann},~\bfnm{Allan}\binits{A.}}
(\byear{2006}).
\btitle{Persistence in forecasting performance and conditional combination
  strategies}.
\bjournal{Journal of Econometrics}
\bvolume{135}
\bpages{31--53}.
\end{barticle}
\endbibitem

\bibitem[\protect\citeauthoryear{Avramov}{2002}]{avramov2002stock}
\begin{barticle}[author]
\bauthor{\bsnm{Avramov},~\bfnm{Doron}\binits{D.}}
(\byear{2002}).
\btitle{Stock return predictability and model uncertainty}.
\bjournal{Journal of Financial Economics}
\bvolume{64}
\bpages{423--458}.
\end{barticle}
\endbibitem

\bibitem[\protect\citeauthoryear{Berry and Stangl}{1996}]{berry1996bayesian}
\begin{barticle}[author]
\bauthor{\bsnm{Berry},~\bfnm{Donald~A}\binits{D.~A.}} \AND
  \bauthor{\bsnm{Stangl},~\bfnm{Dalene~K}\binits{D.~K.}}
(\byear{1996}).
\btitle{Bayesian methods in health-related research}.
\bjournal{Bayesian Biostatistics}
\bpages{3--66}.
\end{barticle}
\endbibitem

\bibitem[\protect\citeauthoryear{Berry et~al.}{2010}]{berry2010bayesian}
\begin{bbook}[author]
\bauthor{\bsnm{Berry},~\bfnm{Scott~M}\binits{S.~M.}},
  \bauthor{\bsnm{Carlin},~\bfnm{Bradley~P}\binits{B.~P.}},
  \bauthor{\bsnm{Lee},~\bfnm{J~Jack}\binits{J.~J.}} \AND
  \bauthor{\bsnm{Muller},~\bfnm{Peter}\binits{P.}}
(\byear{2010}).
\btitle{Bayesian adaptive methods for clinical trials}
\bvolume{38}.
\bpublisher{CRC press}.
\end{bbook}
\endbibitem

\bibitem[\protect\citeauthoryear{Brian}{2010}]{brian2010bayesian}
\begin{barticle}[author]
\bauthor{\bsnm{Brian},~\bfnm{Neelon}\binits{N.}}
(\byear{2010}).
\btitle{Bayesian analysis using power priors with application to pediatric
  quality of care}.
\bjournal{Journal of Biometrics \& Biostatistics}.
\end{barticle}
\endbibitem

\bibitem[\protect\citeauthoryear{Carhart}{1997}]{carhart1997persistence}
\begin{barticle}[author]
\bauthor{\bsnm{Carhart},~\bfnm{Mark~M}\binits{M.~M.}}
(\byear{1997}).
\btitle{On persistence in mutual fund performance}.
\bjournal{The Journal of finance}
\bvolume{52}
\bpages{57--82}.
\end{barticle}
\endbibitem

\bibitem[\protect\citeauthoryear{Carter and Kohn}{1994}]{CarKoh94}
\begin{barticle}[author]
\bauthor{\bsnm{Carter},~\bfnm{C.~K.}\binits{C.~K.}} \AND
  \bauthor{\bsnm{Kohn},~\bfnm{R.}\binits{R.}}
(\byear{1994}).
\btitle{On Gibbs Sampling for State Space Models}.
\bjournal{Biometrika}
\bvolume{81}
\bpages{541-553}.
\end{barticle}
\endbibitem

\bibitem[\protect\citeauthoryear{Chen and Singpurwalla}{1994}]{chen1994non}
\begin{barticle}[author]
\bauthor{\bsnm{Chen},~\bfnm{Yiping}\binits{Y.}} \AND
  \bauthor{\bsnm{Singpurwalla},~\bfnm{Nozer~D}\binits{N.~D.}}
(\byear{1994}).
\btitle{A non-Gaussian Kalman filter model for tracking software reliability}.
\bjournal{Statistica sinica}
\bvolume{4}
\bpages{535--48}.
\end{barticle}
\endbibitem

\bibitem[\protect\citeauthoryear{Dangl and Halling}{2012}]{dangl2012predictive}
\begin{barticle}[author]
\bauthor{\bsnm{Dangl},~\bfnm{Thomas}\binits{T.}} \AND
  \bauthor{\bsnm{Halling},~\bfnm{Michael}\binits{M.}}
(\byear{2012}).
\btitle{Predictive regressions with time-varying coefficients}.
\bjournal{Journal of Financial Economics}
\bvolume{106}
\bpages{157--181}.
\end{barticle}
\endbibitem

\bibitem[\protect\citeauthoryear{Dawid}{1992}]{dawid1992prequential}
\begin{barticle}[author]
\bauthor{\bsnm{Dawid},~\bfnm{A~Philip}\binits{A.~P.}}
(\byear{1992}).
\btitle{Prequential data analysis}.
\bjournal{Lecture Notes-Monograph Series}
\bpages{113--126}.
\end{barticle}
\endbibitem

\bibitem[\protect\citeauthoryear{Fama and French}{1989}]{fama1989business}
\begin{barticle}[author]
\bauthor{\bsnm{Fama},~\bfnm{Eugene~F}\binits{E.~F.}} \AND
  \bauthor{\bsnm{French},~\bfnm{Kenneth~R}\binits{K.~R.}}
(\byear{1989}).
\btitle{Business conditions and expected returns on stocks and bonds}.
\bjournal{Journal of financial economics}
\bvolume{25}
\bpages{23--49}.
\end{barticle}
\endbibitem

\bibitem[\protect\citeauthoryear{Fama and French}{1993}]{fama1993common}
\begin{barticle}[author]
\bauthor{\bsnm{Fama},~\bfnm{Eugene~F}\binits{E.~F.}} \AND
  \bauthor{\bsnm{French},~\bfnm{Kenneth~R}\binits{K.~R.}}
(\byear{1993}).
\btitle{Common risk factors in the returns on stocks and bonds}.
\bjournal{Journal of financial economics}
\bvolume{33}
\bpages{3--56}.
\end{barticle}
\endbibitem

\bibitem[\protect\citeauthoryear{Figlewski}{1984}]{figlewski1984hedging}
\begin{barticle}[author]
\bauthor{\bsnm{Figlewski},~\bfnm{Stephen}\binits{S.}}
(\byear{1984}).
\btitle{Hedging performance and basis risk in stock index futures}.
\bjournal{The Journal of Finance}
\bvolume{39}
\bpages{657--669}.
\end{barticle}
\endbibitem

\bibitem[\protect\citeauthoryear{Gelfand and Dey}{1994}]{gelfand1994bayesian}
\begin{barticle}[author]
\bauthor{\bsnm{Gelfand},~\bfnm{Alan~E}\binits{A.~E.}} \AND
  \bauthor{\bsnm{Dey},~\bfnm{Dipak~K}\binits{D.~K.}}
(\byear{1994}).
\btitle{Bayesian model choice: asymptotics and exact calculations}.
\bjournal{Journal of the Royal Statistical Society. Series B (Methodological)}
\bpages{501--514}.
\end{barticle}
\endbibitem

\bibitem[\protect\citeauthoryear{Gelman et~al.}{2003}]{gelman2003bayesian}
\begin{bbook}[author]
\bauthor{\bsnm{Gelman},~\bfnm{Andrew}\binits{A.}},
  \bauthor{\bsnm{Carlin},~\bfnm{John~B}\binits{J.~B.}},
  \bauthor{\bsnm{Stern},~\bfnm{Hal~S}\binits{H.~S.}} \AND
  \bauthor{\bsnm{Rubin},~\bfnm{Donald~B}\binits{D.~B.}}
(\byear{2003}).
\btitle{Bayesian data analysis}.
\bpublisher{CRC press}.
\end{bbook}
\endbibitem

\bibitem[\protect\citeauthoryear{Geman and Geman}{1984}]{GemGem84}
\begin{barticle}[author]
\bauthor{\bsnm{Geman},~\bfnm{S.}\binits{S.}} \AND
  \bauthor{\bsnm{Geman},~\bfnm{D.}\binits{D.}}
(\byear{1984}).
\btitle{Stochastic relaxation, {G}ibbs distributions, and the {B}ayesian
  restoration of images}.
\bjournal{IEEE Transaction on Pattern Analysis and Machine Intelligence}
\bvolume{6}
\bpages{721-741}.
\end{barticle}
\endbibitem

\bibitem[\protect\citeauthoryear{Grunwald, Raftery and
  Guttorp}{1993}]{grunwald1993time}
\begin{barticle}[author]
\bauthor{\bsnm{Grunwald},~\bfnm{Gary~K}\binits{G.~K.}},
  \bauthor{\bsnm{Raftery},~\bfnm{Adrian~E}\binits{A.~E.}} \AND
  \bauthor{\bsnm{Guttorp},~\bfnm{Peter}\binits{P.}}
(\byear{1993}).
\btitle{Time series of continuous proportions}.
\bjournal{Journal of the Royal Statistical Society. Series B (Methodological)}
\bpages{103--116}.
\end{barticle}
\endbibitem

\bibitem[\protect\citeauthoryear{Hobbs et~al.}{2011}]{hobbs2011hierarchical}
\begin{barticle}[author]
\bauthor{\bsnm{Hobbs},~\bfnm{Brian~P}\binits{B.~P.}},
  \bauthor{\bsnm{Carlin},~\bfnm{Bradley~P}\binits{B.~P.}},
  \bauthor{\bsnm{Mandrekar},~\bfnm{Sumithra~J}\binits{S.~J.}} \AND
  \bauthor{\bsnm{Sargent},~\bfnm{Daniel~J}\binits{D.~J.}}
(\byear{2011}).
\btitle{Hierarchical commensurate and power prior models for adaptive
  incorporation of historical information in clinical trials}.
\bjournal{Biometrics}
\bvolume{67}
\bpages{1047--1056}.
\end{barticle}
\endbibitem

\bibitem[\protect\citeauthoryear{Hoeting, Raftery and
  Madigan}{2002}]{hoeting2002bayesian}
\begin{barticle}[author]
\bauthor{\bsnm{Hoeting},~\bfnm{Jennifer~A}\binits{J.~A.}},
  \bauthor{\bsnm{Raftery},~\bfnm{Adrian~E}\binits{A.~E.}} \AND
  \bauthor{\bsnm{Madigan},~\bfnm{David}\binits{D.}}
(\byear{2002}).
\btitle{Bayesian variable and transformation selection in linear regression}.
\bjournal{Journal of Computational and Graphical Statistics}
\bvolume{11}
\bpages{485--507}.
\end{barticle}
\endbibitem

\bibitem[\protect\citeauthoryear{Ibrahim and Chen}{2000}]{ibrahim2000power}
\begin{barticle}[author]
\bauthor{\bsnm{Ibrahim},~\bfnm{Joseph~G}\binits{J.~G.}} \AND
  \bauthor{\bsnm{Chen},~\bfnm{Ming-Hui}\binits{M.-H.}}
(\byear{2000}).
\btitle{Power prior distributions for regression models}.
\bjournal{Statistical Science}
\bpages{46--60}.
\end{barticle}
\endbibitem

\bibitem[\protect\citeauthoryear{Kass and Raftery}{1995}]{kass1995bayes}
\begin{barticle}[author]
\bauthor{\bsnm{Kass},~\bfnm{Robert~E}\binits{R.~E.}} \AND
  \bauthor{\bsnm{Raftery},~\bfnm{Adrian~E}\binits{A.~E.}}
(\byear{1995}).
\btitle{Bayes factors}.
\bjournal{Journal of the american statistical association}
\bvolume{90}
\bpages{773--795}.
\end{barticle}
\endbibitem

\bibitem[\protect\citeauthoryear{Lewellen and
  Nagel}{2006}]{lewellen2006conditional}
\begin{barticle}[author]
\bauthor{\bsnm{Lewellen},~\bfnm{Jonathan}\binits{J.}} \AND
  \bauthor{\bsnm{Nagel},~\bfnm{Stefan}\binits{S.}}
(\byear{2006}).
\btitle{The conditional CAPM does not explain asset-pricing anomalies}.
\bjournal{Journal of Financial Economics}
\bvolume{82}
\bpages{289--314}.
\end{barticle}
\endbibitem

\bibitem[\protect\citeauthoryear{Madigan and Raftery}{1994}]{madigan1994model}
\begin{barticle}[author]
\bauthor{\bsnm{Madigan},~\bfnm{David}\binits{D.}} \AND
  \bauthor{\bsnm{Raftery},~\bfnm{Adrian~E}\binits{A.~E.}}
(\byear{1994}).
\btitle{Model selection and accounting for model uncertainty in graphical
  models using Occam's window}.
\bjournal{Journal of the American Statistical Association}
\bvolume{89}
\bpages{1535--1546}.
\end{barticle}
\endbibitem

\bibitem[\protect\citeauthoryear{Paez and
  Gamerman}{2013}]{paez2013hierarchical}
\begin{barticle}[author]
\bauthor{\bsnm{Paez},~\bfnm{Marina~Silva}\binits{M.~S.}} \AND
  \bauthor{\bsnm{Gamerman},~\bfnm{Dani}\binits{D.}}
(\byear{2013}).
\btitle{Hierarchical Dynamic Models}.
\bjournal{The SAGE Handbook of Multilevel Modeling}
\bpages{335}.
\end{barticle}
\endbibitem

\bibitem[\protect\citeauthoryear{Petkova and Zhang}{2005}]{petkova2005value}
\begin{barticle}[author]
\bauthor{\bsnm{Petkova},~\bfnm{Ralitsa}\binits{R.}} \AND
  \bauthor{\bsnm{Zhang},~\bfnm{Lu}\binits{L.}}
(\byear{2005}).
\btitle{Is value riskier than growth?}
\bjournal{Journal of Financial Economics}
\bvolume{78}
\bpages{187--202}.
\end{barticle}
\endbibitem

\bibitem[\protect\citeauthoryear{Petris, Petrone and
  Campagnoli}{2009}]{petris2009dynamic}
\begin{bbook}[author]
\bauthor{\bsnm{Petris},~\bfnm{Giovanni}\binits{G.}},
  \bauthor{\bsnm{Petrone},~\bfnm{Sonia}\binits{S.}} \AND
  \bauthor{\bsnm{Campagnoli},~\bfnm{Patrizia}\binits{P.}}
(\byear{2009}).
\btitle{Dynamic linear models with R}.
\bpublisher{Springer}.
\end{bbook}
\endbibitem

\bibitem[\protect\citeauthoryear{Raftery and
  Zheng}{2003}]{raftery2003discussion}
\begin{barticle}[author]
\bauthor{\bsnm{Raftery},~\bfnm{Adrian~E}\binits{A.~E.}} \AND
  \bauthor{\bsnm{Zheng},~\bfnm{Yingye}\binits{Y.}}
(\byear{2003}).
\btitle{Discussion: Performance of Bayesian model averaging}.
\bjournal{Journal of the American Statistical Association}
\bvolume{98}
\bpages{931--938}.
\end{barticle}
\endbibitem

\bibitem[\protect\citeauthoryear{Rapach, Strauss and
  Zhou}{2009}]{rapach2009out}
\begin{barticle}[author]
\bauthor{\bsnm{Rapach},~\bfnm{David~E}\binits{D.~E.}},
  \bauthor{\bsnm{Strauss},~\bfnm{Jack~K}\binits{J.~K.}} \AND
  \bauthor{\bsnm{Zhou},~\bfnm{Guofu}\binits{G.}}
(\byear{2009}).
\btitle{Out-of-sample equity premium prediction: Combination forecasts and
  links to the real economy}.
\bjournal{Review of Financial Studies}
\bpages{hhp063}.
\end{barticle}
\endbibitem

\bibitem[\protect\citeauthoryear{Shephard}{1994}]{shephard1994local}
\begin{barticle}[author]
\bauthor{\bsnm{Shephard},~\bfnm{Neil}\binits{N.}}
(\byear{1994}).
\btitle{Local scale models: State space alternative to integrated GARCH
  processes}.
\bjournal{Journal of Econometrics}
\bvolume{60}
\bpages{181--202}.
\end{barticle}
\endbibitem

\bibitem[\protect\citeauthoryear{Shiller}{2014}]{shiller2014}
\begin{bmisc}[author]
\bauthor{\bsnm{Shiller},~\bfnm{Robert}\binits{R.}}
(\byear{2014}).
\btitle{Online Data}.
\bnote{http://www.econ.yale.edu/~shiller/data.htm}.
\end{bmisc}
\endbibitem

\bibitem[\protect\citeauthoryear{Smith}{1979}]{smith1979generalization}
\begin{barticle}[author]
\bauthor{\bsnm{Smith},~\bfnm{JQ}\binits{J.}}
(\byear{1979}).
\btitle{A generalization of the Bayesian steady forecasting model}.
\bjournal{Journal of the Royal Statistical Society. Series B (Methodological)}
\bpages{375--387}.
\end{barticle}
\endbibitem

\bibitem[\protect\citeauthoryear{Smith}{1981}]{smith1981multiparameter}
\begin{barticle}[author]
\bauthor{\bsnm{Smith},~\bfnm{JQ}\binits{J.}}
(\byear{1981}).
\btitle{The multiparameter steady model}.
\bjournal{Journal of the Royal Statistical Society. Series B (Methodological)}
\bpages{256--260}.
\end{barticle}
\endbibitem

\bibitem[\protect\citeauthoryear{Stock and Watson}{2004}]{stock2004combination}
\begin{barticle}[author]
\bauthor{\bsnm{Stock},~\bfnm{James~H}\binits{J.~H.}} \AND
  \bauthor{\bsnm{Watson},~\bfnm{Mark~W}\binits{M.~W.}}
(\byear{2004}).
\btitle{Combination forecasts of output growth in a seven-country data set}.
\bjournal{Journal of Forecasting}
\bvolume{23}
\bpages{405--430}.
\end{barticle}
\endbibitem

\bibitem[\protect\citeauthoryear{Tan et~al.}{2002}]{tan2002bayesian}
\begin{barticle}[author]
\bauthor{\bsnm{Tan},~\bfnm{SB}\binits{S.}},
  \bauthor{\bsnm{Machin},~\bfnm{D}\binits{D.}},
  \bauthor{\bsnm{Tai},~\bfnm{BC}\binits{B.}},
  \bauthor{\bsnm{Foo},~\bfnm{KF}\binits{K.}} \AND
  \bauthor{\bsnm{Tan},~\bfnm{EH}\binits{E.}}
(\byear{2002}).
\btitle{A Bayesian re-assessment of two Phase II trials of gemcitabine in
  metastatic nasopharyngeal cancer}.
\bjournal{British journal of cancer}
\bvolume{86}
\bpages{843--850}.
\end{barticle}
\endbibitem

\bibitem[\protect\citeauthoryear{Welch and
  Goyal}{2008}]{welch2008comprehensive}
\begin{barticle}[author]
\bauthor{\bsnm{Welch},~\bfnm{Ivo}\binits{I.}} \AND
  \bauthor{\bsnm{Goyal},~\bfnm{Amit}\binits{A.}}
(\byear{2008}).
\btitle{A comprehensive look at the empirical performance of equity premium
  prediction}.
\bjournal{Review of Financial Studies}
\bvolume{21}
\bpages{1455--1508}.
\end{barticle}
\endbibitem

\bibitem[\protect\citeauthoryear{West and Harrison}{1998}]{west1998bayesian}
\begin{barticle}[author]
\bauthor{\bsnm{West},~\bfnm{M}\binits{M.}} \AND
  \bauthor{\bsnm{Harrison},~\bfnm{J}\binits{J.}}
(\byear{1998}).
\btitle{Bayesian Forecasting and Dynamic Models (2nd edn)}.
\bjournal{Journal of the Operational Research Society}
\bvolume{49}
\bpages{179--179}.
\end{barticle}
\endbibitem

\bibitem[\protect\citeauthoryear{Wilmott}{1995}]{wilmott1995mathematics}
\begin{bbook}[author]
\bauthor{\bsnm{Wilmott},~\bfnm{Paul}\binits{P.}}
(\byear{1995}).
\btitle{The mathematics of financial derivatives: a student introduction}.
\bpublisher{Cambridge University Press}.
\end{bbook}
\endbibitem

\end{thebibliography}

\end{document}